\documentclass[journal]{IEEEtran}

\usepackage[dvips]{color}
\usepackage{graphicx}
\usepackage{amsmath}
\begin{document}
%
\title{Optimal Relay Selection for Physical-Layer Security in Cooperative Wireless Networks}
%
%
%

\markboth{IEEE Journal on Selected Areas in Communications (accepted to appear)}%
{Yulong Zou \MakeLowercase{\textit{et al.}}: Optimal Relay Selection
for Physical-Layer Security in Cooperative Wireless Networks}

\author{Yulong~Zou,~Xianbin~Wang, and Weiming~Shen
\thanks{Copyright (c) 2013 IEEE. Personal use of this material is permitted. However, permission to
use this material for any other purposes must be obtained from the
IEEE by sending a request to pubs-permissions@ieee.org.}
\thanks{Y. Zou is with the Electrical and Computer Engineering Department, University of
Western Ontario, London, ON N6A 5B9, Canada, and also with the
Institute of Signal Processing and Transmission, Nanjing University
of Posts and Telecommunications, Nanjing 210003, China. E-mail:
yulong.zou@gmail.com.}
\thanks{X. Wang is with the Electrical and Computer Engineering Department, University of Western
Ontario, London, ON N6A 5B9, Canada.}
\thanks{W. Shen is with the College of Electronics and Information Engineering,
Tongji University, Shanghai, China.}
\thanks{This work was partially supported by the Auto21 Network of Centre of Excellence, Canada,
and the National Natural Science Foundation of China (Grant No.
61271240).}

}

\maketitle

\begin{abstract}
In this paper, we explore the physical-layer security in cooperative
wireless networks with multiple relays where both
amplify-and-forward (AF) and decode-and-forward (DF) protocols are
considered. We propose the AF and DF based optimal relay selection
(i.e., AF\emph{b}ORS and DF\emph{b}ORS) schemes to improve the
wireless security against eavesdropping attack. For the purpose of
comparison, we examine the traditional AF\emph{b}ORS and
DF\emph{b}ORS schemes, denoted by T-AF\emph{b}ORS and
T-DF\emph{b}ORS, respectively. {We also investigate a so-called
multiple relay combining (MRC) framework and present the traditional
AF and DF based MRC schemes, called T-AF\emph{b}MRC and
T-DF\emph{b}MRC, where multiple relays participate in forwarding the
source signal to destination which then combines its received
signals from the multiple relays.} We derive closed-form intercept
probability expressions of the proposed AF\emph{b}ORS and
DF\emph{b}ORS (i.e., P-AF\emph{b}ORS and P-DF\emph{b}ORS) as well as
the T-AF\emph{b}ORS, T-DF\emph{b}ORS, T-AF\emph{b}MRC and
T-DF\emph{b}MRC schemes in the presence of eavesdropping attack. We
further conduct an asymptotic intercept probability analysis to
evaluate the diversity order performance of relay selection schemes
and show that no matter which relaying protocol is considered (i.e.,
AF and DF), the traditional and proposed optimal relay selection
approaches both achieve the diversity order $M$ where $M$ represents
the number of relays. In addition, numerical results show that for
both AF and DF protocols, the intercept probability performance of
proposed optimal relay selection is strictly better than that of the
traditional relay selection and multiple relay combining methods.

\end{abstract}


\IEEEpeerreviewmaketitle

\section{Introduction}
%
%
%
%
\IEEEPARstart {M}{ultiple-input} multiple-output (MIMO) [1], [2] has
been widely recognized as an effective way to combat wireless fading
and increase link throughput by exploiting multiple antennas at both
the transmitter and receiver. However, it may be difficult to
implement multiple antennas in some cases (e.g., handheld terminals,
sensor nodes, etc.) due to the limitation in physical size and power
consumption. As an alternative, user cooperation [3] is now emerging
as a promising paradigm to achieve the spatial diversity by enabling
user terminals to share their antennas and form a virtual antenna
array. Until recently, there has been extensive research on the user
cooperation from different perspectives, e.g., cooperative resource
allocation [4], performance analysis and optimization [5], [6], and
cooperative medium access control (MAC) and routing design [7], [8].

User cooperation not only improves the reliability and throughput of
wireless transmissions, but also has great potential to enhance the
wireless security against eavesdropping attack. Differing from the
conventional encryption techniques relying on secret keys,
physical-layer security exploits the physical characteristics of
wireless channels to prevent the eavesdropper from intercepting the
signal transmission from source to its intended destination. It has
been proven in [9] and [10] that in the presence of an eavesdropper,
a so-called \emph{secrecy capacity} is shown as the difference
between the channel capacity from source to destination (called main
link) and that from source to eavesdropper (called wiretap link).
Moreover, if the secrecy capacity is negative, the eavesdropper will
succeed in intercepting the source signal and an intercept event
occurs in this case. However, due to the fading effect, the secrecy
capacity is severely limited in wireless communications. To that
end, user cooperation as an emerging spatial diversity technique can
effectively combat wireless fading and thus improves the secrecy
capacity of wireless transmissions in the presence of eavesdropping
attack.

At present, most of existing work on the user cooperation for
wireless security is focused on developing the secrecy capacity from
an information-theoretic perspective. In [11], the authors studied
the secrecy capacity of wireless transmissions in the presence of an
eavesdropper with a relay node, where the amplify-and-forward (AF),
decode-and-forward (DF), and compress-and-forward (CF) relaying
protocols are examined and compared with each other. The cooperative
jamming was proposed in [12] and analyzed in terms of the achievable
secrecy rate, where multiple users are allowed to cooperate with
each other in preventing eavesdropping attack. In [13], the
cooperation strategy was further examined to enhance the
physical-layer security and a so-called noise-forwarding scheme was
proposed, where the relay node attempts to send codewords
independent of the source message to confuse the eavesdropper. In
[14] and [15], the authors studied the cooperative relays for
enhancing physical-layer security and showed the secrecy capacity
improvement by using cooperative relays. {The physical-layer
security was further examined in two-way relay networks in [16] and
[17] where multiple two-way relays are exploited to improve the
secrecy capacity against eavesdropping attack. In addition, the
authors of [18] and [19] investigated the physical-layer security in
MIMO relay networks and showed the significant improvement in terms
of secrecy capacity through the use of MIMO relays.}

In this paper, we consider a cooperative wireless network with
multiple relays in the presence of an eavesdropper and examine the
optimal relay selection to improve physical-layer security against
eavesdropping attack. Differing from the traditional relay selection
in [20]-[22] where only the channel state information (CSI) of
two-hop relay links (i.e., source-relay and relay-destination) are
considered, we here have to take into account additional CSI of the
wiretap links, in addition to the two-hop relay links' CSI. The main
contributions of this paper are summarized as follows. {Firstly},
considering AF and DF relaying protocols, we propose the AF and DF
based optimal relay selection schemes which are denoted by
P-AF{\it{b}}ORS and P-DF{\it{b}}ORS, respectively. We also examine
the traditional AF and DF based optimal relay selection (i.e.,
T-AF{\it{b}}ORS and T-DF{\it{b}}ORS) and multiple relay combining
(i.e., T-AF{\it{b}}MRC and T-DF{\it{b}}MRC) as benchmark schemes.
Secondly, we derive closed-form expressions of intercept probability
for the P-AF{\it{b}}ORS and P-DF{\it{b}}ORS as well as the
T-AF{\it{b}}ORS, T-DF{\it{b}}ORS, T-AF{\it{b}}MRC and
T-DF{\it{b}}MRC schemes in Rayleigh fading channels. It is shown
that for both AF and DF protocols, the intercept probability of
proposed optimal relay selection is always smaller than that of the
traditional relay selection and multiple relay combining approaches,
which shows the advantage of proposed optimal relay selection.
Finally, we evaluate the diversity order performance of optimal
relay selection schemes and show that no matter which relaying
protocol is considered, the proposed and traditional optimal relay
selection schemes both achieve the same diversity order $M$, where
$M$ represents the number of relays.

The remainder of this paper is organized as follows. Section II
presents the system model and proposes the conventional direct
transmission, T-AF\emph{b}ORS, T-DF\emph{b}ORS, T-AF\emph{b}MRC,
T-DF\emph{b}MRC, P-AF\emph{b}ORS, and P-DF\emph{b}ORS schemes. In
Section III, we derive closed-form intercept probability expressions
of the direct transmission, T-AF\emph{b}ORS, T-DF\emph{b}ORS,
T-AF\emph{b}MRC, T-DF\emph{b}MRC, P-AF\emph{b}ORS, and
P-DF\emph{b}ORS schemes in the presence of eavesdropping attack. In
Section IV, we analyze the diversity order performance of the
traditional and proposed relay selection schemes. Next, in Section
V, numerical evaluation is conducted to show the advantage of
proposed optimal relay selection over traditional relay selection
and multiple relay combining approaches in terms of the intercept
probability. Finally, we make some concluding remarks in Section VI.

\section{System Model and Proposed Optimal Relay Selection Schemes}

\subsection{System Model}
\begin{figure}
  \centering
  {\includegraphics[scale=0.65]{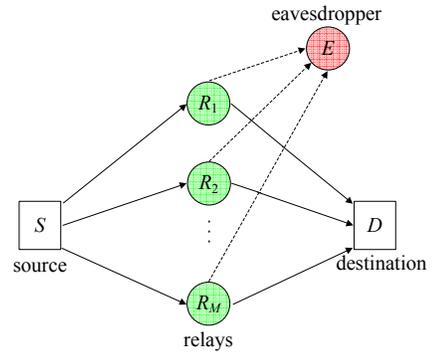}\\
  \caption{A cooperative wireless network consisting of one source, one destination, and multiple relays in the presence of an eavesdropper.}\label{Fig1}}
\end{figure}

Consider a cooperative wireless network consisting of one source,
one destination, and $M$ relays in the presence of an eavesdropper
as shown in Fig. 1, where all nodes are equipped with single antenna
and the solid and dash lines represent the main and wiretap links,
respectively. The main and wiretap links both are modeled as
Rayleigh fading channels and the thermal noise received at any node
is modeled as a complex Gaussian random variable with zero mean and
variance $\sigma^2_n$, i.e., ${\mathcal{CN}}(0,\sigma^2_n)$.
Following [14], we consider that $M$ relays are exploited to assist
the transmission from source to destination and the direct links
from source to destination and eavesdropper are not available, e.g.,
the destination and eavesdropper both are out of the coverage area.
For notational convenience, $M$ relays are denoted by ${\cal{R}} =
\{ {{R}}_i |i = 1,2, \cdots ,M\}$. Differing from the existing work
[14] in which all relays participate in forwarding the source
messages to destination, we here consider the use of the optimal
relay only to assist the message transmission from source to
destination. More specifically, the source node first broadcasts the
message to cooperative relays among which only the best relay will
be selected to forward its received signal to destination by using
either amplify-and-forward (AF) or decode-and-forward (DF)
strategies. Meanwhile, the eavesdropper monitors the transmission
from the optimal relay to destination and attempts to interpret the
source message. Following [11] and [14], we assume that the
eavesdropper knows everything about the signal transmission from
source via relay to destination, including the encoding scheme at
source, forwarding protocol at relays, and decoding method at
destination, except that the source signal is confidential.

It is pointed out that in order to effectively prevent the
eavesdropper from intercepting, the optimal relay selection not only
has to consider the CSI of main links to maximize the channel
capacity from source to destination, but also needs to take into
account the wiretap links' CSI to minimize the channel capacity from
source to eavesdropper. This differs from the traditional relay
selection in [20]-[22] where only the two-hop relay links' CSI is
considered in performing the best relay selection. Similarly to [14]
and [23], we here assume that the global CSI of both main and
wiretap links is available, which is a common assumption in the
physical-layer security literature. Notice that the wiretaps link's
CSI can be estimated and obtained by monitoring the eavesdropper's
transmissions as discussed in [23]. {Moreover, if the eavesdropper's
CSI is unknown, we can consider the use of traditional relay
selection [20]-[22] which does not require the CSI of wiretap
links.} In the following, we first present the conventional direct
transmission without relay as a benchmark scheme and then propose
the AF and DF based optimal relay selection schemes to improve the
physical-layer security against eavesdropping attack.
\subsection{Direct Transmission}
For comparison purpose, this subsection describes the conventional
direct transmission without relay. Consider that the source
transmits a signal $s$ ($E(|s{|^2}) = 1$) with power $P$. Thus, the
received signal at destination is expressed as
\begin{equation}\label{equa1}
r_d=\sqrt{P}h_{sd}s+n_d,
\end{equation}
where $h_{sd}$ represents a fading coefficient of the channel from
source to destination and $n_d \sim {\mathcal{CN}}(0,\sigma^2_n)$
represents additive white Gaussian noise (AWGN) at destination.
{Notice that the channel coefficient $h_{sd}$ is modeled as Rayleigh
fading which corresponds to an ideal OFDM subchannel [24] and [25].}
Meanwhile, due to the broadcast nature of wireless transmissions,
the eavesdropper also receives a copy of the source signal $s$ and
the corresponding received signal is written as
\begin{equation}\label{equa2}
r_e=\sqrt{P}h_{se}s+n_e,
\end{equation}
where $h_{se}$ represents a fading coefficient of the channel from
source to eavesdropper and $n_e \sim {\mathcal{CN}}(0,\sigma^2_n)$
represents AWGN at eavesdropper. Assuming the optimal Gaussian
codebook used at source, the maximal achievable rate (also known as
channel capacity) of the direct transmission from source to
destination is obtained from Eq. (1) as
\begin{equation}\label{equa3}
C^{\textrm{direct}}_{sd}=\log_2(1+\dfrac{|h_{sd}|^2P}{\sigma^2_n}),
\end{equation}
where ${\sigma^2_n}$ is the noise variance. Similarly, from Eq. (2),
the capacity of wiretap link from source to eavesdropper is easily
given by
\begin{equation}\label{equa4}
C^{\textrm{direct}}_{se}=\log_2(1+\dfrac{|h_{se}|^2P}{\sigma^2_n}).
\end{equation}
It has been proven in [10] that the secrecy capacity is shown as the
difference between the capacity of main link and that of wiretap
link. Hence, the secrecy capacity of direct transmission is given by
\begin{equation}\label{equa5}
C^{\textrm{direct}}_{s}=C^{\textrm{direct}}_{sd} - C^{\textrm{direct}}_{se},
\end{equation}
where $C^{\textrm{direct}}_{sd}$ and $C^{\textrm{direct}}_{se}$ are
given in Eqs. (3) and (4), respectively. As discussed in [10], when
the secrecy capacity is negative (i.e., the capacity of main link
falls below the wiretap link's capacity), the eavesdropper will
succeed in intercepting the source signal and an intercept event
occurs. Thus, the probability that the eavesdropper successfully
intercepts source signal, called \emph{intercept probability}, is a
key metric in evaluating the performance of physical-layer security.
In this paper, we mainly focus on how to improve the intercept
probability by exploiting cooperative relays for the physical-layer
security enhancement. The following subsections propose the optimal
relay selection by considering AF and DF protocols, respectively.
\subsection{Amplify-and-Forward}
In this subsection, we consider the AF relaying protocol in which
the relay will forward a scaled version of its received source
signal to destination without any sort of decoding. To be specific,
the source node first broadcasts the signal $s$ to $M$ relays. Then,
the optimal relay node will be selected to transmit a scaled version
of its received signal. Notice that in the AF relaying process, the
source signal $s$ is transmitted twice from the source and relay. In
order to make a fair comparison with the direct transmission, the
total amount of transmit power at source and relay shall be limited
to $P$. By using the equal-power allocation for simplicity, the
transmit power at source and relay is given by $P/2$. Thus,
considering that the source node transmits its signal $s$ with power
$P/2$, the received signal at relay $R_i$ can be given by
\begin{equation}\label{equa6}
r_i=\sqrt {\frac{P}{2}} h_{si}s+n_i,
\end{equation}
where $h_{si}$ represents a fading coefficient of the channel from
source to $R_i$ and $n_i \sim {\mathcal{CN}}(0,\sigma^2_n)$
represents AWGN at $R_i$. Without loss of generality, consider that
$R_i$ is selected as the optimal relay to forward its received
signal to destination. Assuming that the CSI $h_{si}$ is available,
$R_i$ first performs coherent detection by multiplying $r_i$ with
$h^*_{si}$ and then normalizes $h^*_{si}r_i$ with a scaling factor
$\frac{1}{{|{h_{si}}{|^2}\sqrt {P/2} }}$. After that, $R_i$
transmits the normalized $h^*_{si}r_i$ with power $P/2$ to
destination, thus the received signal at destination is given by
\begin{equation}\label{equa7}
\begin{split}
{r_d} &= \sqrt {\frac{P}{2}} {h_{id}}\frac{{{h^*_{si}r_i}}}{{|{h_{si}}{|^2}\sqrt {P/2} }} + {n_d}\\
&=  \sqrt {\frac{P}{2}} {h_{id}}s + \frac{{{h_{id}}h_{si}^*}}{{|{h_{si}}{|^2} }}{n_i} + {n_d},
\end{split}
\end{equation}
from which the capacity of AF relaying transmission from $R_i$ to destination is given by
\begin{equation}\label{equa8}
C^{\textrm{AF}}_{id}=\log_2\left(1+\frac{{|{h_{si}}{|^2}|{h_{id}}{|^2}P}}{{2(|{h_{si}}{|^2} + |{h_{id}}{|^2})\sigma _n^2}}\right).
\end{equation}
Meanwhile, the received signal at eavesdropper from $R_i$ is expressed as
\begin{equation}\label{equa9}
{r_e} =  \sqrt {\frac{P}{2}} {h_{ie}}s + \frac{{{h_{ie}}h_{si}^*}}{{|{h_{si}}{|^2} }}{n_i} + {n_e}.
\end{equation}
Similarly to Eq. (8), we obtain the capacity of AF relaying transmission from $R_i$ to eavesdropper as
\begin{equation}\label{equa10}
C^{\textrm{AF}}_{ie}=\log_2\left(1+\frac{{|{h_{si}}{|^2}|{h_{ie}}{|^2}P}}{{2(|{h_{si}}{|^2} + |{h_{ie}}{|^2})\sigma _n^2}}\right).
\end{equation}
Combining Eqs. (8) and (10), we can easily obtain the secrecy capacity of AF relaying transmission with $R_i$ as
\begin{equation}\label{equa11}
\begin{split}
C^{\textrm{AF}}_{i}&=C^{\textrm{AF}}_{id} - C^{\textrm{AF}}_{ie}\\
&={\log _2}\left( {\dfrac{{1 + \dfrac{{|{h_{si}}{|^2}|{h_{id}}{|^2}P}}
{{2(|{h_{si}}{|^2} + |{h_{id}}{|^2})\sigma _n^2}}}}{{1 + \dfrac{{|{h_{si}}
{|^2}|{h_{ie}}{|^2}P}}{{2(|{h_{si}}{|^2} + |{h_{ie}}{|^2})\sigma _n^2}}}}} \right).
\end{split}
\end{equation}
Next, we discuss how to determine the optimal relay and propose the
AF based optimal relay selection scheme denoted by P-AF{\it{b}}ORS
for notational convenience. For the comparison purpose, the
traditional AF based optimal relay selection and multiple relay
combining (i.e., T-AF{\it{b}}ORS and T-AF\emph{b}MRC) are also
presented.
\subsubsection{P-AFbORS}Now, let us consider the P-AF{\it{b}}ORS scheme in which the relay
that maximizes the secrecy capacity of AF relaying transmission is
viewed as the optimal relay. Thus, the AF based optimal relay
selection criterion can be obtained from Eq. (11) as
\begin{equation}\label{equa12}
\begin{split}
\textrm{OptimalRelay} &= \arg \mathop {\max }\limits_{i \in {\cal{R}}}C^{\textrm{AF}}_{i}\\
&= \arg \mathop {\max }\limits_{i \in {\cal{R}}} {\dfrac{{1 + \dfrac{{|{h_{si}}{|^2}|{h_{id}}{|^2}P}}
{{2(|{h_{si}}{|^2} + |{h_{id}}{|^2})\sigma _n^2}}}}{{1 + \dfrac{{|{h_{si}}{|^2}|{h_{ie}}{|^2}P}}{{2(|{h_{si}}{|^2}
 + |{h_{ie}}{|^2})\sigma _n^2}}}}},
\end{split}
\end{equation}
where ${\cal{R}}$ represents a set of $M$ relays. One can observe
from Eq. (12) that the P-AF{\it{b}}ORS scheme takes into account not
only the main links' CSI $|h_{si}|^2$ and $|h_{id}|^2$, but also the
wiretap link's CSI $|h_{ie}|^2$. Notice that the transmit power $P$
in Eq. (12) is a known parameter and the noise variance $\sigma
_n^2$ is shown as $\sigma _n^2 =\kappa TB$ [26], where $\kappa $ is
Boltzmann constant (i.e., $\kappa  = 1.38 \times {10^{ - 23}}$), $T$
is room temperature, and $B$ is system bandwidth. Since the room
temperature $T$ and system bandwidth $B$ both are predetermined, the
noise variance $\sigma_n^2$ can be easily obtained. It is pointed
out that using the proposed optimal relay selection criterion in Eq.
(12), we can further develop a centralized or distributed relay
selection algorithm. To be specific, for a centralized relay
selection, the source node needs to maintain a table that consists
of $M$ relays and related CSI (i.e., $|h_{si}|^2$, $|h_{id}|^2$ and
$|h_{ie}|^2$). In this way, the optimal relay can be easily
determined by looking up the table using the proposed criterion in
Eq. (12), which is referred to as centralized relay selection
strategy. For a distributed relay selection, each relay maintains a
timer and sets an initial value of the timer in inverse proportional
to $[{{1 + \frac{{|{h_{si}}{|^2}|{h_{id}}{|^2}P}}{{2(|{h_{si}}{|^2}
+ |{h_{id}}{|^2})\sigma _n^2}}}}]/[{{1 +
\frac{{|{h_{si}}{|^2}|{h_{ie}}{|^2}P}}{{2(|{h_{si}}{|^2} +
|{h_{ie}}{|^2})\sigma _n^2}}}}] $, resulting in the optimal relay
with the smallest initial value for its timer. As a consequence, the
optimal relay exhausts its timer earliest compared with the other
relays, and then broadcasts a control packet to notify the source
node and other relays [21].
\subsubsection{T-AFbORS}For the purpose of comparison, we here present the traditional AF
based optimal relay selection (T-AF{\it{b}}ORS) scheme. Since the
wiretap link's CSI $|h_{ie}|^2$ is not considered in T-AF{\it{b}}ORS
scheme, the relay with the largest $C^{\textrm{AF}}_{id}$ (i.e., the
capacity of AF relaying transmission from $R_i$ to destination) is
selected as the optimal relay. Therefore, the traditional AF based
optimal relay selection criterion is obtained from Eq. (8) as
\begin{equation}\label{equa13}
\begin{split}
\textrm{OptimalRelay} &= \arg \mathop {\max }\limits_{i \in {\cal{R}}}C^{\textrm{AF}}_{id}\\
&= \arg \mathop {\max }\limits_{i \in {\cal{R}}}\dfrac{{|{h_{si}}{|^2}|{h_{id}}{|^2}}}{{|{h_{si}}{|^2}
+ |{h_{id}}{|^2}}},
\end{split}
\end{equation}
which is the traditional harmonic mean policy as given by Eq. (2) in
[20]. It is shown from Eq. (13) that only the main links' CSI
$|h_{si}|^2$ and $|h_{id}|^2$ is taken into account in the
T-AF{\it{b}}ORS scheme, differing from the P-AF{\it{b}}ORS scheme
that requires the CSI of both main and wiretap links (i.e.,
$|h_{si}|^2$, $|h_{id}|^2$ and $|h_{ie}|^2$).
{\subsubsection{T-AFbMRC}This subsection presents the traditional AF
based multiple relay combining (T-AF{\it{b}}MRC) scheme, where all
AF relays participate in forwarding the source signal transmission
to destination which combines its received signals from the multiple
AF relays. Notice that in the T-AF{\it{b}}MRC scheme, the total
amount of transmit power consumed at the source and $M$ relays
should be constrained to a fixed value (i.e., $P$). With the
equal-power allocation, the transmit power for each node (e.g., the
source and relays) is given by $P/(M+1)$. Thus, the source node
first transmits the signal $s$ with power $P/(M+1)$ to $M$ relays
that will normalize their received signals with respective scaling
factors $\frac{1}{{|{h_{si}}{|^2}\sqrt {P/(M+1)} }}$ wherein $i =
1,2, \cdots ,M$. Then, all relays forward their normalized signals
to destination with power $P/(M+1)$. Hence, the received signal at
destination from relay $R_i$ can be expressed as
\begin{equation}\label{equa14}
{r^i_d} =  \sqrt {\frac{P}{M+1}} {h_{id}}s + \frac{{{h_{id}}h_{si}^*}}{{|{h_{si}}{|^2} }}{n_i} + {n^i_d},
\end{equation}
where ${n_i}$ and ${n^i_d}$ represent AWGN received at relay $R_i$
and destination, respectively. The destination combines its received
signals from multiple AF relays, where the combining coefficient
$|h_{si}|^2h^{*}_{id}$ is considered for the received signal
${r^i_d}$ from relay $R_i$. Accordingly, the combined signal denoted
by $r_d$ at destination is given by
\begin{equation}\label{equa15}
\begin{split}
{r_d} =& \sum\limits_{i = 1}^M {\sqrt {\frac{P}{{M + 1}}} |{h_{si}}{|^2}|{h_{id}}{|^2}s} \\
& + \sum\limits_{i = 1}^M {\left( {|{h_{id}}{|^2}h_{si}^*{n_i} + |{h_{si}}{|^2}h_{id}^*{n^i_d}} \right)},
\end{split}
\end{equation}
from which the transmission capacity from source to destination via
$M$ relays with the T-AF{\it{b}}MRC scheme is given by
\begin{equation}\label{equa16}
C_{sd}^{{\textrm{AF}}b{\textrm{MRC}}} = {\log _2}\left( {1 + \frac{{{(
 \sum\limits_{i = 1}^M {|{h_{si}}{|^2}|{h_{id}}{|^2}}  )^2}P}}{{(M + 1)
 \sum\limits_{i = 1}^M {H(h_{si},h_{id}) \sigma _n^2} }}} \right),
\end{equation}
where $H(h_{si},h_{id}) = {|{h_{si}}{|^2}|{h_{id}}{|^4} + |{h_{si}}
 {|^4}|{h_{id}}{|^2}}$ and $\sigma^2_n$ represents the noise variance. Also, the
transmission capacity from source to eavesdropper with the
T-AF{\it{b}}MRC scheme is similarly obtained as
\begin{equation}\label{equa17}
C_{se}^{{\textrm{AF}}b{\textrm{MRC}}} = {\log _2}\left( {1 + \frac{{{( \sum\limits_{i = 1}^M
 {|{h_{si}}{|^2}|{h_{ie}}{|^2}}  )^2}P}}{{(M + 1)\sum\limits_{i = 1}
 ^M {H(h_{si},h_{ie})\sigma _n^2} }}} \right),
\end{equation}
where $H(h_{si},h_{ie})={|{h_{si}}{|^2} |{h_{ie}}{|^4} +
|{h_{si}}{|^4}|{h_{ie}}{|^2}}$. Hence, the secrecy capacity of
T-AF{\it{b}}MRC scheme is shown as
\begin{equation}\label{equa18}
C_{s}^{{\textrm{AF}}b{\textrm{MRC}}} = C_{sd}^{{\textrm{AF}}b{\textrm{MRC}}}
- C_{se}^{{\textrm{AF}}b{\textrm{MRC}}} ,
\end{equation}
where $C_{sd}^{{\textrm{AF}}b{\textrm{MRC}}} $ and
$C_{se}^{{\textrm{AF}}b{\textrm{MRC}}}$ are given by Eqs. (16) and
(17), respectively.}
\subsection{Decode-and-Forward}
This subsection mainly focuses on the DF relaying protocol in which
the relay first decodes its received signal from source and then
re-encodes and transmits its decoded outcome to the destination.
More specifically, the source node first broadcasts the signal $s$
to $M$ relays that attempt to decode their received signals. Then,
only the optimal relay is selected to re-encode and transmit its
decoded outcome to the destination. Similarly to AF relaying
protocol, the total transmit power at source and relay with DF
protocol is also limited to $P$ in order to make a fair comparison
with the direct transmission. Considering the equal-power
allocation, we obtain the transmit power at source and relay as
$P/2$. It has been shown in [2] that the capacity of DF relaying
transmission is the minimum of the capacity from source to relay and
that from relay to destination, since either source-relay or
relay-destination links in failure will result in the two-hop DF
transmission in failure. Hence, considering $R_i$ as the optimal
relay, we can obtain the capacity of DF transmission from source via
$R_i$ to destination as
\begin{equation}\label{equa19}
C^{\textrm{DF}}_{sid}=\min({C_{si},C_{id}}),
\end{equation}
where $C_{si}$ and $C_{id}$, respectively, represent the channel
capacity from source to $R_i$ and that from $R_i$ to destination,
which are given by
\begin{equation}\label{equa20}
C_{si}=\log_2(1+\dfrac{|h_{si}|^2P}{2\sigma^2_n}),
\end{equation}
and
\begin{equation}\label{equa21}
C_{id}=\log_2(1+\dfrac{|h_{id}|^2P}{2\sigma^2_n}).
\end{equation}
Meanwhile, the eavesdropper can overhear the transmission from $R_i$
to destination. Hence, the channel capacity from $R_i$ to
eavesdropper can be easily obtained as
\begin{equation}\label{equa22}
C^{\textrm{DF}}_{ie}=\log_2(1+\dfrac{|h_{ie}|^2P}{2\sigma^2_n}).
\end{equation}
Combining Eqs. (19) and (22), the secrecy capacity of DF relaying transmission with $R_i$ is given by
\begin{equation}\label{equa23}
\begin{split}
C^{\textrm{DF}}_{i}&=C^{\textrm{DF}}_{sid} - C^{\textrm{DF}}_{ie}\\
&={\log _2}\left(1 + \frac{{\min ( |{h_{si}}{|^2},|{h_{id}}{|^2}) P}}{{2\sigma _n^2}}\right)\\
&\quad - {\log _2}\left(1 + \frac{{|{h_{ie}}{|^2}P}}{{2\sigma _n^2}}\right).
\end{split}
\end{equation}
In the following subsections, we present the P-DF{\it{b}}ORS and
T-DF{\it{b}}ORS schemes, respectively. For the comparison purpose,
the traditional DF based multiple relay combining (T-DF{\it{b}}MRC)
scheme is also discussed.
\subsubsection{P-DFbORS}Let us first consider P-DF{\it{b}}ORS scheme. Similarly to
P-AF{\it{b}}ORS scheme, we consider the relay that maximizes the
secrecy capacity of DF relaying transmission as the optimal relay.
Thus, the DF based optimal relay selection criterion is easily
obtained from Eq. (23) as
\begin{equation}\label{equa24}
\begin{split}
\textrm{OptimalRelay} &= \arg \mathop {\max }\limits_{i \in {\cal{R}}}C^{\textrm{DF}}_{i}\\
&=\arg \mathop {\max }\limits_{i \in {\cal{R}}} \frac{{\min ( |{h_{si}}{|^2},|{h_{id}}{|^2}) P
+ 2\sigma _n^2}}{{|{h_{ie}}{|^2}P + 2\sigma _n^2}},
\end{split}
\end{equation}
which shows that the global CSI of both main and wiretap links
(i.e., $|h_{si}|^2$, $|h_{id}|^2$ and $|h_{ie}|^2$) is required in
determining the optimal relay.
\subsubsection{T-DFbORS}We now present the traditional DF based optimal relay selection
(T-DF{\it{b}}ORS) scheme in which the relay that maximizes the
capacity of DF relaying transmission $C^{\textrm{DF}}_{sid}$ is
selected as the optimal relay. Thus, the traditional DF based
optimal relay selection criterion is obtained from Eq. (19) as
\begin{equation}\label{equa25}
\begin{split}
\textrm{OptimalRelay} &= \arg \mathop {\max }\limits_{i \in {\cal{R}}}C^{\textrm{DF}}_{sid}\\
&= \arg \mathop {\max }\limits_{i \in {\cal{R}}}\min({|h_{si}|^2,|h_{id}|^2}),
\end{split}
\end{equation}
which is the traditional max-min relay selection criterion as given
by Eq. (1) in [20]. As shown in Eq. (25), only the main links' CSI
$|h_{si}|^2$ and $|h_{id}|^2$ is taken into account in
T-DF{\it{b}}ORS scheme without considering the wiretap link's CSI
$|h_{ie}|^2$. {\subsubsection{T-DFbMRC}This subsection presents the
T-DF\emph{b}MRC scheme where multiple DF relays will assist the
signal transmission from source to destination. To be specific, the
source node first transmits its signal $s$ with power $P/2$ to $M$
relays which then attempt to decode their received signals. For
notational convenience, these relays that succeed in decoding the
source signal are represented by a set $ D$, called \emph{decoding
set}, where the sample space of decoding set is given by $\Omega  =
\{ D|D \in \emptyset  \cup {D_m},m = 1,2, \cdots ,{2^M} - 1\} $,
where $\cup$ denotes the union operation, $\emptyset$ denotes empty
set, and $D_m$ denotes a non-empty subcollection of $M$ relays. If
the decoding set is empty (i.e., all relays fail to decode the
source signal), no relay will transmit and thus both the destination
and eavesdropper can not interpret the source signal. If the
decoding set $D$ is not empty (i.e., $D=D_m$), all relays in $D_m$
are selected to forward their decoded outcomes to destination, where
the total transmit power of multiple relays in the decoding set is
constrained to $P/2$. With the equal-power allocation, the transmit
power for each relay in decoding set $D_m$ is given by $P/|D_m|$,
where $|D_m|$ represents the cardinality of set $D_m$ (i.e., the
number of elements in set $D_m$). Thus, considering that relay $R_i
\in D_m$ transmits its decoded result $s$ with power $P/|D_m|$, the
received signal at destination is given by
\begin{equation}\label{equa26}
{r^i_d} =  \sqrt {\frac{P}{|D_m|}} {h_{id}}s + {n^i_d}.
\end{equation}
Then, the destination combines its received signals from multiple DF
relays in decoding set $D_m$ with the maximal ratio combining. Thus,
the combined signal denoted by $r_d$ at destination can be written
as
\begin{equation}\label{equa27}
{r_d} = \sum\limits_{i \in {D_m}} {\sqrt {\frac{P}{|D_m|}}|{h_{id}}{|^2}s}
+ \sum\limits_{i \in {D_m}} {h_{id}^*n_d^i} ,
\end{equation}
from which the transmission capacity from source to destination with
the T-DF{\it{b}}MRC scheme in the case of $D=D_m$ is given by
\begin{equation}\label{equa28}
C_{sd}^{{\textrm{DF}}b{\textrm{MRC}}} \left( {D = {D_m}} \right)= {\log _2}\left( {1
+ \sum\limits_{i \in {D_m}} {\frac{|{h_{id}}{|^2}P}{{|{D_m}|\sigma _n^2}}} } \right),
\end{equation}
where $\sigma_n^2$ represents the noise variance. Similarly, the
transmission capacity from source to eavesdropper with the
T-DF{\it{b}}MRC scheme can be obtained as
\begin{equation}\label{equa29}
C_{se}^{{\textrm{DF}}b{\textrm{MRC}}} \left( {D = {D_m}} \right)=
 {\log _2}\left( {1 + \sum\limits_{i \in {D_m}} {\frac{|{h_{ie}}{|^2}P}{{|{D_m}|\sigma _n^2}}} } \right).
\end{equation}
Hence, combining Eqs. (28) and (29), the secrecy capacity of
T-DF{\it{b}}MRC scheme in the case of $D=D_m$ is given by
\begin{equation}\label{equa30}
\begin{split}
C_{s}^{{\textrm{DF}}b{\textrm{MRC}}}\left( {D = {D_m}} \right) =& C_{sd}^{{\textrm{DF}}b{\textrm{MRC}}}
 \left( {D = {D_m}} \right)\\
&- C_{se}^{{\textrm{DF}}b{\textrm{MRC}}}\left( {D = {D_m}} \right),
\end{split}
\end{equation}
which completes the signal modeling of T-DF{\it{b}}MRC scheme.
}
\section{Intercept Probability Analysis over Rayleigh Fading Channels}
In this section, we derive closed-form intercept probability
expressions of conventional direct transmission, P-AF{\it{b}}ORS,
P-DF{\it{b}}ORS, T-AF{\it{b}}ORS, T-DF{\it{b}}ORS, T-AF{\it{b}}MRC,
and T-DF{\it{b}}MRC schemes over Rayleigh fading channels.
\subsection{Direct Transmission}
Let us first analyze the intercept probability of direct
transmission as a baseline for comparison purpose. As is known, an
intercept event occurs when the secrecy capacity becomes negative.
Thus, the intercept probability of direct transmission is obtained
from Eq. (5) as
\begin{equation}\label{equa31}
\begin{split}
P_{{\textrm{intercept}}}^{\textrm{direct}}& = \Pr \left( {{C^{\textrm{direct}}_{sd}}
< {C^{\textrm{direct}}_{se}}} \right) \\
&= \Pr \left( {|{h_{sd}}{|^2} < |{h_{se}}{|^2}} \right),
\end{split}
\end{equation}
where the second equation is obtained by using Eqs. (3) and (4).
Since the Rayleigh fading model is used throughout this paper, we
can obtain that $|h_{sd}|^2$ and $|h_{se}|^2$ follow exponential
distributions. Thus, a closed-form intercept probability expression
of direct transmission is given by
\begin{equation}\label{equa32}
P_{{\textrm{intercept}}}^{\textrm{direct}} = \dfrac{{\sigma _{se}^2}}{{\sigma _{se}^2 + \sigma _{sd}^2}},
\end{equation}
where $\sigma _{se}^2 = E(|{h_{se}}{|^2})$ and $\sigma _{sd}^2 =
E(|{h_{sd}}{|^2})$. It is observed from Eq. (32) that the intercept
probability of direct transmission is independent of the transmit
power $P$, which implies that the wireless security performance
cannot be improved by increasing the transmit power. This also
motivates us to exploit cooperative relays to decrease the intercept
probability and improve the physical-layer security.
\subsection{P-AFbORS}
In this subsection, we present the intercept probability analysis of
P-AF{\it{b}}ORS scheme. Considering the fact that an intercept event
occurs when the secrecy capacity falls below zero, we can obtain the
intercept probability of P-AF{\it{b}}ORS scheme from Eq. (12) as
\begin{equation}\label{equa33}
\begin{split}
P_{{\textrm{intercept}}}^{\textrm{P-AF{\it{b}}ORS}} &= \Pr \left( {\mathop {\max }
\limits_{i \in {\mathcal{R}}} C_i^{\textrm{AF}} < 0} \right) \\
&= \prod\limits_{i = 1}^M {\Pr \left( {|{h_{ie}}{|^2} > |{h_{id}}{|^2}} \right)} ,
\end{split}
\end{equation}
where the second equation is obtained by using Eq. (11). Considering
that $|h_{ie}|^2$ and $|h_{id}|^2$ are independent exponentially
distributed random variables, we obtain
\begin{equation}\label{equa34}
P_{{\textrm{intercept}}}^{\textrm{P-AF{\it{b}}ORS}}  = \prod\limits_{i = 1}^M
{\dfrac{{\sigma _{ie}^2}}{{\sigma _{ie}^2 + \sigma _{id}^2}}} ,
\end{equation}
where $\sigma _{ie}^2 = E(|{h_{ie}}{|^2})$ and $\sigma _{id}^2 = E(|{h_{id}}{|^2})$.
\subsection{P-DFbORS}
This subsection derives a closed-form intercept probability
expression of P-DF{\it{b}}ORS scheme. According to the definition of
intercept event, an intercept probability of P-DF{\it{b}}ORS scheme
is obtained from Eq. (24) as
\begin{equation}\label{equa35}
\begin{split}
P_{{\textrm{intercept}}}^{\textrm{P-DF{\it{b}}ORS}} &= \Pr \left( {\mathop {\max }\limits_{i \in {\mathcal{R}}} C_i^{\textrm{DF}} < 0} \right) \\
&=\prod\limits_{i = 1}^M {\Pr \left\{ {\min (|{h_{si}}{|^2},|{h_{id}}{|^2}) < |{h_{ie}}{|^2}} \right\}}  ,
\end{split}
\end{equation}
where the second equation is obtained by using Eq. (23). Notice that
random variables $|{h_{si}}{|^2}$, $|{h_{id}}{|^2}$ and
$|{h_{ie}}{|^2}$ follow exponential distributions with means
$\sigma^2_{si}$, $\sigma^2_{id}$ and $\sigma^2_{ie}$, respectively.
Denoting $X=\min (|{h_{si}}{|^2}, |{h_{id}}{|^2})$, we can easily
obtain the cumulative density function (CDF) of $X$ as
\begin{equation}\label{equa36}
{P_X}(X < x) = 1 - \exp ( - \frac{x}{{\sigma _{si}^2}} - \frac{x}{{\sigma _{id}^2}}),
\end{equation}
wherein $x\ge0$. Using Eq. (36), we have
\begin{equation}\label{equa37}
\begin{split}
&{\Pr \left\{ {\min (|{h_{si}}{|^2},|{h_{id}}{|^2}) < |{h_{ie}}{|^2}} \right\}} \\
&= \int_0^\infty  {[1 - \exp ( - \dfrac{x}{{\sigma _{si}^2}} - \dfrac{x}{{\sigma _{id}^2}})]\dfrac{1}{{\sigma _{ie}^2}}\exp ( - \dfrac{x}{{\sigma _{ie}^2}})dx} \\
&= \dfrac{{\sigma _{id}^2\sigma _{ie}^2 + \sigma _{si}^2\sigma _{ie}^2}}{{\sigma _{id}^2\sigma _{ie}^2 + \sigma _{si}^2\sigma _{ie}^2 + \sigma _{si}^2\sigma _{id}^2}}.
\end{split}
\end{equation}
Substituting Eq. (37) into Eq. (35) gives
\begin{equation}\label{equa38}
P_{{\textrm{intercept}}}^{\textrm{P-DF{\it{b}}ORS}}  = \prod\limits_{i = 1}^M
 \dfrac{{\sigma _{id}^2\sigma _{ie}^2 + \sigma _{si}^2\sigma _{ie}^2}}{{\sigma _{id}
 ^2\sigma _{ie}^2 + \sigma _{si}^2\sigma _{ie}^2 + \sigma _{si}^2\sigma _{id}^2}}.
\end{equation}
In addition, we can easily prove $\frac{{\sigma _{ie}^2}}{{\sigma
_{ie}^2 + \sigma _{si}^2}} < \frac{{\sigma _{id}^2\sigma _{ie}^2 +
\sigma _{si}^2\sigma _{ie}^2}}{{\sigma _{id}^2\sigma _{ie}^2 +
\sigma _{si}^2\sigma _{ie}^2 + \sigma _{si}^2\sigma _{id}^2}}$.
Considering $\frac{{\sigma _{ie}^2}}{{\sigma _{ie}^2 + \sigma
_{si}^2}}>0$ and $ \frac{{\sigma _{id}^2\sigma _{ie}^2 + \sigma
_{si}^2\sigma _{ie}^2}}{{\sigma _{id}^2\sigma _{ie}^2 + \sigma
_{si}^2\sigma _{ie}^2 + \sigma _{si}^2\sigma _{id}^2}}>0$, we obtain
\begin{equation}\label{equa39}
\prod\limits_{i = 1}^M {\dfrac{{\sigma _{ie}^2}}{{\sigma _{ie}^2 + \sigma _{si}^2}}}
 < \prod\limits_{i = 1}^M {\dfrac{{\sigma _{id}^2\sigma _{ie}^2 + \sigma _{si}^2
 \sigma _{ie}^2}}{{\sigma _{id}^2\sigma _{ie}^2 + \sigma _{si}^2\sigma _{ie}^2
 + \sigma _{si}^2\sigma _{id}^2}}},
\end{equation}
which theoretically shows that the intercept probability of
P-AF{\it{b}}ORS scheme is strictly less than that of P-DF{\it{b}}ORS
scheme, implying the advantage of AF relaying protocol over DF
protocol from the physical-layer security perspective.
\subsection{T-AFbORS}
In this subsection, we present the intercept probability analysis of
T-AF{\it{b}}ORS scheme. From Eq. (13), we obtain an intercept
probability of T-AF{\it{b}}ORS scheme as
\begin{equation}\label{equa40}
P_{{\textrm{intercept}}}^{\textrm{T-AF{\it{b}}ORS}} = \Pr \left( {\mathop {\max }\limits_{i \in {\cal{R}}} C_{id}^{\textrm{AF}} < C_{oe}^{\textrm{AF}}} \right),
\end{equation}
where $C_{oe}^{\textrm{AF}}$ denotes the channel capacity from the
optimal relay to eavesdropper. It is pointed out that the
T-AF{\it{b}}ORS scheme does not consider the eavesdropper's CSI
$|h_{ie}|^2$. This means that the traditional relay selection is
independent of the eavesdropper's channel information. Using the law
of total probability, the intercept probability of T-AF{\it{b}}ORS
scheme is given by
\begin{equation}\label{equa41}
\begin{split}
P_{{\textrm{intercept}}}^{\textrm{T-AF{\it{b}}ORS}} = \sum\limits_{m = 1}^M &{\Pr \left( {{\textrm{OptimalRelay}} = m} \right)}\\
&\times{\Pr \left( {\mathop {\max }\limits_{i \in {\cal{R}}} C_{id}^{\textrm{AF}} < C_{me}^{\textrm{AF}}} \right)}.
\end{split}
\end{equation}
For simplicity, we here consider that fading coefficients ${h_{si}}$
and ${h_{id}}{\textrm{ }}(i=1, \cdots ,M)$ are identically and
independently distributed, leading to $\Pr \left(
{{\textrm{OptimalRelay}} = m} \right) =1/{M}$. Substituting this
result and Eq. (13) into Eq. (41) yields
\begin{equation}\label{equa42}
\begin{split}
P_{{\textrm{intercept}}}^{\textrm{T-AF{\it{b}}ORS}} =
\sum\limits_{m = 1}^M {\frac{1}{M}\Pr \left( \begin{split} &{\mathop {\max }\limits_{i \in {\cal{R}}}
 \frac{{|{h_{si}}{|^2}|{h_{id}}{|^2}}}{{|{h_{si}}{|^2} + |{h_{id}}{|^2}}}}\\
 &< {\frac{{|{h_{sm}}{|^2}|{h_{me}}{|^2}}}{{|{h_{sm}}{|^2} + |{h_{me}}{|^2}}}} \end{split} \right)}.
\end{split}
\end{equation}
It is noted that obtaining a closed-form solution to Eq. (42) is
challenging, however numerical intercept probability results of
T-AF{\it{b}}ORS scheme can be easily obtained through computer
simulations.

\subsection{T-DFbORS}
This subsection analyzes the intercept probability of
T-DF{\it{b}}ORS scheme in Rayleigh fading channels. From Eq. (25),
we obtain an intercept probability of T-DF{\it{b}}ORS scheme as
\begin{equation}\label{equa43}
P_{{\textrm{intercept}}}^{\textrm{T-DF{\it{b}}ORS}} = \Pr \left( {\mathop {\max }
\limits_{i \in {\cal{R}}} C_{sid}^{\textrm{DF}} < C_{oe}^{\textrm{DF}}} \right),
\end{equation}
where $C_{oe}^{\textrm{DF}}$ denotes the channel capacity from the
optimal relay to eavesdropper with DF relaying protocol. Similarly,
assuming that ${h_{si}}$ and ${h_{id}}{\textrm{ }}(i=1, \cdots ,M)$
are identically and independently distributed and using the law of
total probability, the intercept probability of T-DF{\it{b}}ORS
scheme is given by
\begin{equation}\label{equa44}
P_{{\textrm{intercept}}}^{\textrm{T-DF{\it{b}}ORS}} = \sum\limits_{m = 1}^M
{\frac{1}{M}\Pr \left( \begin{split} &{\mathop {\max }\limits_{i \in {\cal{R}}} \min
\left( {|{h_{si}}{|^2},|{h_{id}}{|^2}} \right)} \\
&< {|{h_{me}}{|^2}} \end{split} \right)} .
\end{equation}
Notice that $|{h_{si}}{|^2}$, $|{h_{id}}{|^2}$ and $|{h_{me}}{|^2}$
follow exponential distributions with means $\sigma^2_{si}$,
$\sigma^2_{id}$ and $\sigma^2_{me}$, respectively. Letting
$x=|{h_{me}}{|^2}$, we obtain Eq. (45) at the top of following page,
\begin{figure*}
\begin{equation}\label{equa45}
\begin{split}
P_{{\textrm{intercept}}}^{\textrm{T-DF{\it{b}}ORS}}&= \sum\limits_{m = 1}^M
{\frac{1}{M}\int_0^\infty  {\prod\limits_{i = 1}^M {[1 - \exp ( - \frac{x}
{{\sigma _{si}^2}} - \frac{x}{{\sigma _{id}^2}})]} \frac{1}{{\sigma _{me}^2}}
\exp ( - \frac{x}{{\sigma _{me}^2}})dx} } \\
&= \sum\limits_{m = 1}^M {\frac{1}{M}\int_0^\infty  {\left( {1 + \sum\limits_{k = 1}^{{2^M} - 1}
{{{( - 1)}^{|{{\mathcal{A}}_k}|}}\exp [ - \sum\limits_{i \in {{\mathcal{A}}_k}}
 {(\frac{x}{{\sigma _{si}^2}} + \frac{x}{{\sigma _{id}^2}})} ]} } \right)\frac{1}
 {{\sigma _{me}^2}}\exp ( - \frac{x}{{\sigma _{me}^2}})dx} } \\
&= \sum\limits_{m = 1}^M {\frac{1}{M}\left( {1 + \sum\limits_{k = 1}^{{2^M} - 1}
 {{{( - 1)}^{|{{\mathcal{A}}_k}|}}{[1 + \sum\limits_{i \in {{\mathcal{A}}_k}}
  {(\frac{{\sigma _{me}^2}}{{\sigma _{si}^2}} + \frac{{\sigma _{me}^2}}
  {{\sigma _{id}^2}})} ]^{- 1}}} } \right)}
\end{split}
\end{equation}
\end{figure*}
where the second equation is obtained by using the binomial
expansion, ${\mathcal{A}}_k$ represents the $k$-th non-empty
sub-collection of $M$ relays, and $|{\mathcal{A}}_k|$ represents the
number of elements in set ${\mathcal{A}}_k$. {\subsection{T-AFbMRC}
This subsection presents the intercept probability analysis of
T-AF{\it{b}}MRC scheme. From Eq. (18), an intercept probability of
the T-AF{\it{b}}MRC scheme is obtained as
\begin{equation}\label{equa46}
P_{{\textrm{intercept}}}^{\textrm{T-AF{\it{b}}MRC}}
= \Pr \left( C_{sd}^{{\textrm{AF}}b{\textrm{MRC}}} < C_{se}
^{{\textrm{AF}}b{\textrm{MRC}}} \right).
\end{equation}
Substituting Eqs. (16) and (17) into Eq. (46) gives
\begin{equation}\label{equa47}
P_{{\textrm{intercept}}}^{\textrm{T-AF{\it{b}}MRC}}
= \Pr \left( \begin{split} &{\frac{{{(\sum\limits_{i = 1}^M {|{h_{si}}{|^2}|{h_{id}}
{|^2}} )^2}}}{{\sum\limits_{i = 1}^M {H(h_{si},h_{id})} }}} \\
&< {\frac{{{(\sum\limits_{i = 1}^M {|{h_{si}}{|^2}
|{h_{ie}}{|^2}} )^2}}}{{\sum\limits_{i = 1}^M {H(h_{si},h_{ie})} }}} \end{split} \right),
\end{equation}
where $H(h_{si},h_{id}) = {|{h_{si}}{|^2}|{h_{id}}{|^4} + |{h_{si}}
 {|^4}|{h_{id}}{|^2}}$ and $H(h_{si},h_{ie}) = {|{h_{si}}{|^2}|{h_{ie}}{|^4} + |{h_{si}}
 {|^4}|{h_{ie}}{|^2}}$. From Eq. (47), the numerical intercept probability results of
T-AF\emph{b}MRC scheme can be easily determined through computer
simulations.} {\subsection{T-DFbMRC} In this subsection, the
intercept probability analysis of T-DF{\it{b}}MRC scheme is
presented. Using the law of total probability, we can obtain an
intercept probability of the T-DF{\it{b}}MRC scheme from Eq. (30) as
\begin{equation}\label{equa48}
\begin{split}
P_{{\textrm{intercept}}}^{\textrm{T-DF{\it{b}}MRC}} =&
 \sum\limits_{m = 1}^{{2^M} - 1} {\Pr \left( {D = {D_m}} \right)}\\
 &\quad \quad \times {\Pr  \left( {C_s^{\textrm{DF{\it{b}}MRC}}(D = {D_m}) < 0} \right)},
\end{split}
\end{equation}
where $\Pr \left( {D = {D_m}} \right)$ represents the probability of
occurrence of event $D=D_m$. Notice that if the decoding set is
empty, all relays keep silent and nothing is transmitted, implying
that the eavesdropper can not intercept the source signal.
Substituting Eqs. (28) and (29) into Eq. (48)
yields
\begin{equation}\label{equa49}
\begin{split}
P_{{\textrm{intercept}}}^{\textrm{T-DF{\it{b}}MRC}} =& \sum\limits_{m
= 1}^{{2^M} - 1} {\Pr \left( {D = {D_m}} \right)} \\
&\quad \quad \times {\Pr \left({\sum\limits_{i \in {D_m}} {|{h_{id}}{|^2}}  < \sum\limits_{i \in
{D_m}} {|{h_{ie}}{|^2}} } \right)}.
\end{split}
\end{equation}
According to Shannon's channel coding theorem, relay $R_i$ can
succeed in decoding the source signal if no outage event occurs over
the channel from source to relay $R_i$. Otherwise, relay $R_i$ is
deemed to fail to decode the source signal. Thus, the probability of
occurrence of event $D=D_m$ can be given by
\begin{equation}\label{equa50}
\Pr \left( {D = {D_m}} \right) = \prod\limits_{i \in {D_m}}
{(1 - {\rm{Pou}}{{\rm{t}}_i})} \prod\limits_{i \in {{\bar D}_m}}
{{\rm{Pou}}{{\rm{t}}_i}},
\end{equation}
where $\bar D_m  = {\cal{R}} - D_m$ represents the complementary set
of $D_m $ and ${\rm{Pou}}{{\rm{t}}_i}$ represents the probability of
occurrence of outage event over the channel from source to relay
$R_i$. So far, we have completed the intercept probability analysis
of direct transmission, P-AF{\it{b}}ORS, P-DF{\it{b}}ORS,
T-AF{\it{b}}ORS, T-DF{\it{b}}ORS, T-AF{\it{b}}MRC, and
T-DF{\it{b}}MRC schemes.}
\section{Diversity Order Analysis}
In this section, we analyze the diversity order performance of the
traditional and proposed optimal relay selection schemes including
the T-AF{\it{b}}ORS, T-DF{\it{b}}ORS, P-AF{\it{b}}ORS, and
P-DF{\it{b}}ORS. Let us first recall the traditional definition of
diversity gain. As shown in [27], the traditional diversity gain is
given by {\begin{equation}\label{equa51} d =  - \mathop {\lim
}\limits_{{\textrm{SNR}} \to \infty } \dfrac{\log
{P_e}({\textrm{SNR}})}{\log {\textrm{SNR}}},
\end{equation}}
where ${\textrm{SNR}}$ denotes signal-to-noise ratio (SNR) and
$P_e({\textrm{SNR}})$ denotes bit error rate. One can observe from
the preceding equation that the traditional diversity gain is
defined as ${\textrm{SNR}} \to \infty$. However, the intercept
probability expressions as shown in Eqs. (32), (34) and (38) are
independent of SNR, resulting in that the traditional diversity gain
definition is not applicable here. To that end, we propose a
generalized diversity gain as follows
\begin{equation}\label{equa52}
{d_{{\textrm{generalized}}}} = -\mathop {\lim }\limits_{{\lambda
_{de}} \to \infty } \dfrac{{\log (P_{{\textrm{intercept}}})}}{{\log
({\lambda _{de}})}},
\end{equation}
where $\lambda_{de}=\sigma^2_{sd}/\sigma^2_{se}$ is the ratio of
average channel gain from source to destination to that from source
to eavesdropper, which is referred to as the main-to-eavesdropper
ratio (MER) throughout this paper.
\subsection{Direct Transmission}
Let us first analyze the diversity order of direct transmission as a
baseline. From Eqs. (32) and (52), the diversity order of direct
transmission is obtained as
\begin{equation}\label{equa53}
{d_{{\textrm{direct}}}} = -\mathop {\lim }\limits_{{\lambda _{de}}
\to \infty } \dfrac{{\log
(P_{{\textrm{intercept}}}^{{\textrm{direct}}})}}{{\log ({\lambda
_{de}})}}=1,
\end{equation}
which shows that the direct transmission achieves the diversity
order of only one, i.e., the intercept probability of direct
transmission scheme behaves as $\frac{1}{\lambda_{de}}$ in high
{main-to-eavesdropper ratio} (MER) regions.
\subsection{P-AFbORS}
This subsection presents the diversity order analysis of
P-AF{\it{b}}ORS scheme. Similarly, the diversity order of
P-AF{\it{b}}ORS scheme is given by
\begin{equation}\label{equa54}
{d_{{\textrm{P-AF{\it{b}}ORS}}}} = -\mathop {\lim }\limits_{{\lambda
_{de}} \to \infty } \dfrac{{\log
(P_{{\textrm{intercept}}}^{{\textrm{P-AF{\it{b}}ORS}}})}}{{\log
({\lambda _{de}})}},
\end{equation}
where $P_{{\textrm{intercept}}}^{{\textrm{P-AF{\it{b}}ORS}}}$ is
given in Eq. (34). Denoting $\sigma _{id}^2 = {\alpha _{id}}\sigma
_{sd}^2$ and $\sigma _{ie}^2 = {\alpha _{ie}}\sigma _{se}^2$, we can
rewrite $P_{{\textrm{intercept}}}^{{\textrm{P-AF{\it{b}}ORS}}}$ from
Eq. (34) as
\begin{equation}\label{equa55}
P_{{\textrm{intercept}}}^{{\textrm{P-AF{\it{b}}ORS}}}=\prod\limits_{i = 1}^M
{\dfrac{{{\alpha _{ie}}}}{{{\alpha _{ie}}\lambda _{de}^{ - 1} + {\alpha _{id}}}}}
  \cdot {(\frac{1}{{{\lambda _{de}}}})^M},
\end{equation}
where $\lambda_{de}=\sigma^2_{sd}/\sigma^2_{se}$. Substituting Eq. (55) into Eq. (54) gives
\begin{equation}\label{equa56}
{d_{{\textrm{P-AF{\it{b}}ORS}}}} =M,
\end{equation}
which shows that the diversity order $M$ is achieved by
P-AF{\it{b}}ORS scheme. One can see that as the number of relays $M
$ increases, the diversity order of P-AF{\it{b}}ORS scheme increases
accordingly, showing that increasing the number of relays can
significantly improve the intercept probability
performance.
\subsection{P-DFbORS} In this subsection, we focus on
the diversity order analysis of P-DF{\it{b}}ORS scheme. Similarly to
Eq. (54), the diversity order of P-DF{\it{b}}ORS scheme is given by
\begin{equation}\label{equa57}
{d_{{\textrm{P-DF{\it{b}}ORS}}}} = -\mathop {\lim }\limits_{{\lambda
_{de}} \to \infty } \dfrac{{\log
(P_{{\textrm{intercept}}}^{{\textrm{P-DF{\it{b}}ORS}}})}}{{\log
({\lambda _{de}})}},
\end{equation}
where $P_{{\textrm{intercept}}}^{{\textrm{P-DF{\it{b}}ORS}}}$ is
given in Eq. (38). Denoting $\sigma _{id}^2 = {\alpha _{id}}\sigma
_{sd}^2$, $\sigma _{ie}^2 = {\alpha _{ie}}\sigma _{se}^2$ and
$\sigma _{si}^2 = {\alpha _{si}}\sigma _{sd}^2$, we can rewrite
$P_{{\textrm{intercept}}}^{{\textrm{P-DF{\it{b}}ORS}}}$ from Eq.
(38) as
\begin{equation}\label{equa58}
\begin{split}
P_{{\textrm{intercept}}}^{{\textrm{P-DF{\it{b}}ORS}}}=&\prod\limits_{i = 1}^M
{\dfrac{{{\alpha _{id}} + {\alpha _{si}}}}{{{\alpha _{id}}\lambda _{de}^{ - 1}
+ {\alpha _{si}}\lambda _{de}^{ - 1} + {\alpha _{si}}{\alpha _{id}}{\alpha^{-1}
_{ie}}}}} \\
& \quad \times {(\frac{1}{{{\lambda _{de}}}})^M},
\end{split}
\end{equation}
where $\lambda_{de}=\sigma^2_{sd}/\sigma^2_{se}$. Substituting Eq. (58) into Eq. (57) yields
\begin{equation}\label{equa59}
{d_{{\textrm{P-DF{\it{b}}ORS}}}} =M.
\end{equation}
It is shown from Eq. (59) that the P-DF{\it{b}}ORS scheme achieves
the diversity order $M$, i.e., the intercept probability of
P-DF{\it{b}}ORS scheme behaves as $(\frac{1}{\lambda_{de}})^M$ for
${\lambda_{de}} \to \infty$.
\subsection{T-AFbORS} We now examine the
diversity order of T-AF{\it{b}}ORS scheme. The diversity order of
T-AF{\it{b}}ORS scheme is given by
\begin{equation}\label{equa60}
{d_{{\textrm{T-AF{\it{b}}ORS}}}} = -\mathop {\lim }\limits_{{\lambda _{de}}
\to \infty } \dfrac{{\log (P_{{\textrm{intercept}}}^
{{\textrm{T-AF{\it{b}}ORS}}})}}{{\log ({\lambda _{de}})}},
\end{equation}
where $P_{{\textrm{intercept}}}^{{\textrm{T-AF{\it{b}}ORS}}}$ is
given in Eq. (42). Denoting $X = |{h_{sm}}{|^2}$ and $Y =
|{h_{me}}{|^2}$ and using the conditional probability, we obtain Eq.
(61),
\begin{figure*}
\begin{equation}\label{equa61}
P_{{\textrm{intercept}}}^{{\textrm{T-AF{\it{b}}ORS}}}
= \sum\limits_{m = 1}^M {\frac{1}{M}\int_0^\infty {\int_0^\infty
 {\prod\limits_{i = 1,i \ne m}^M {\Pr \left( {\frac{{|{h_{si}}{|^2}
 |{h_{id}}{|^2}}}{{|{h_{si}}{|^2} + |{h_{id}}{|^2}}} < \frac{{xy}}
 {{x + y}}} \right)} \Pr \left( {|{h_{md}}{|^2} < y} \right)f(x,y)dxdy} } }
\end{equation}
\end{figure*}
where $f(x,y)$ represents a joint probability density function (PDF)
of $(X,Y)$. Considering that $X$ and $Y$ are independent
exponentially distributed, the joint PDF $f(x,y)$ is given by
\begin{equation}\label{equa62}
f(x,y) = \dfrac{1}{{\sigma _{sm}^2\sigma _{me}^2}}
\exp ( - \dfrac{x}{{\sigma _{sm}^2}} - \dfrac{y}{{\sigma _{me}^2}}),
\end{equation}
where $\sigma _{sm}^2 = E(|{h_{sm}}{|^2})$ and $\sigma _{me}^2 =
E(|{h_{me}}{|^2})$. Using inequalities $\frac{1}{{|{h_{si}}{|^2}}} +
\frac{1}{{|{h_{id}}{|^2}}} \ge \max
(\frac{1}{{|{h_{si}}{|^2}}},\frac{1}{{|{h_{id}}{|^2}}})$ and
$\frac{1}{x} + \frac{1}{y} \le 2\max (\frac{1}{x},\frac{1}{y})$, we
obtain
\begin{equation}\label{equa63}
\begin{split}
&\Pr \left( {\frac{{|{h_{si}}{|^2}|{h_{id}}{|^2}}}{{|{h_{si}}{|^2}
+ |{h_{id}}{|^2}}} < \frac{{xy}}{{x + y}}} \right) \\
&= \Pr \left( {\frac{1}{{|{h_{si}}{|^2}}} + \frac{1}{{|{h_{id}}{|^2}}}
> \frac{1}{x} + \frac{1}{y}} \right)\\
&\ge \Pr \left( {\max (\frac{1}{{|{h_{si}}{|^2}}},\frac{1}
{{|{h_{id}}{|^2}}}) > 2\max (\frac{1}{x},\frac{1}{y})} \right)\\
&= \Pr \left( {\min (|{h_{si}}{|^2},|{h_{id}}{|^2}) < \frac{1}{2}\min (x,y)} \right)\\
&= 1 - \exp [( - \frac{1}{{2\sigma _{si}^2}} - \frac{1}{{2\sigma _{id}^2}})\min (x,y)].
\end{split}
\end{equation}
Substituting $\Pr \left(
{\frac{{|{h_{si}}{|^2}|{h_{id}}{|^2}}}{{|{h_{si}}{|^2} +
|{h_{id}}{|^2}}} < \frac{{xy}}{{x + y}}} \right) \ge 1 - \exp [( -
\frac{1}{{2\sigma _{si}^2}} - \frac{1}{{2\sigma _{id}^2}})\min
(x,y)]$ from Eq. (63) into Eq. (61), we can obtain a lower bound on
the intercept probability of T-AF{\it{b}}ORS scheme as
\begin{equation}\label{equa64}
\begin{split}
P_{{\textrm{intercept}}}^{{\textrm{T-AF{\it{b}}ORS}}} \ge & \sum\limits_{m = 1}^M
 {\frac{1}{M}\int_0^\infty  {\int_0^\infty { \prod\limits_{i = 1,i \ne m}^M  g_i (x,y)}}}\\
 &\quad \quad \times {[1 - \exp ( - \frac{y}{{\sigma _{md}^2 }})]f(x,y) }  dxdy,
\end{split}
\end{equation}
where $g_i (x,y) = 1 - \exp [( - \frac{1}{{2\sigma _{si}^2}} -
\frac{1}{{2\sigma _{id}^2}})\min (x,y)]$ and $f(x,y)$ is given by
Eq. (62). {\textbf{Proposition 1:} \emph{Given independent
exponential random variables $x$ and $y$ with respective means
$\sigma _{sm}^2$ and $\sigma _{me}^2$, the following equations
hold,}
\begin{equation}\nonumber
1 - \exp [( - \frac{1}{{2\sigma _{si}^2}} - \frac{1}{{2\sigma _{id}^2}})
\min (x,y)] = (\frac{1}{{2\sigma _{si}^2}} + \frac{1}{{2\sigma _{id}^2}})\min (x,y),
\end{equation}
and
\begin{equation}\nonumber
1 - \exp ( - \frac{y}{{\sigma _{md}^2}}) = \frac{y}{{\sigma _{md}^2}},
\end{equation}
\emph{for ${\lambda _{de}} \to \infty $, where $\lambda_{de}=\sigma^2_{sd}/\sigma^2_{se}$.}\\
\textbf{Proof: }\emph{See Appendix A for details.}}\\
Using Proposition 1 and denoting $\sigma _{si}^2 = {\alpha
_{si}}\sigma _{sd}^2$, $\sigma _{id}^2 = {\alpha _{id}}\sigma
_{sd}^2$, $\sigma _{sm}^2 = {\alpha _{sm}}\sigma _{sd}^2$, $\sigma
_{md}^2 = {\alpha _{md}}\sigma _{sd}^2$ and $\sigma _{me}^2 =
{\alpha _{me}}\sigma _{se}^2$, we obtain from Eq. (64) as Eq. (65)
with ${\lambda _{de}} \to \infty $ at the top of following page.
\begin{figure*}
\begin{equation}\label{equa65}
\begin{split}
P_{{\textrm{intercept}}}^{{\textrm{T-AF{\it{b}}ORS}}} &\ge \sum\limits_{m = 1}^M
 {\frac{1}{M}\prod\limits_{i = 1,i \ne m}^M {(\frac{1}{{2\sigma _{si}^2}} +
 \frac{1}{{2\sigma _{id}^2}})} \int_0^\infty  {\int_0^\infty  {\frac{{{{[\min (x,y)]}
 ^{M - 1}}y}}{{\sigma _{sm}^2\sigma _{md}^2\sigma _{me}^2}}\exp ( - \frac{x}
 {{\sigma _{sm}^2}} - \frac{y}{{\sigma _{me}^2}})dxdy} } } \\
&= \sum\limits_{m = 1}^M {\frac{1}{M}\prod\limits_{i = 1,i \ne m}^M {(\frac{1}
{{2\sigma _{si}^2}} + \frac{1}{{2\sigma _{id}^2}})} \frac{1}{{\sigma _{md}^2}}
\int_0^\infty  {\frac{x^{M - 1}}{\sigma _{sm}^2}\exp ( - \frac{x}{{\sigma _{sm}^2}})
dx\int_x^\infty  {\frac{y}{\sigma _{me}^2}\exp ( - \frac{y}{{\sigma _{me}^2}})dy} } } \\
&\quad+ \sum\limits_{m = 1}^M {\frac{1}{M}\prod\limits_{i = 1,i \ne m}^M {(\frac{1}
{{2\sigma _{si}^2}} + \frac{1}{{2\sigma _{id}^2}})} \frac{1}{{\sigma _{md}^2}}
\int_0^\infty  {{\frac{y^M}{\sigma _{me}^2}}\exp ( - \frac{y}{{\sigma _{me}^2}})
dy\int_y^\infty  {\frac{1}{\sigma _{sm}^2}\exp ( - \frac{x}{{\sigma _{sm}^2}})dx} } } \\
&= \sum\limits_{m = 1}^M {\frac{1}{M}\prod\limits_{i = 1,i \ne m}^M {(\frac{M+1}
{2{{\alpha _{si}}}} + \frac{M+1}{2{{\alpha _{id}}}})} \frac{{(M - 1)!\alpha _{me}
^{M + 1}}}{{{\alpha _{sm}}{\alpha _{md}}}}}  \cdot {(\frac{1}{{{\lambda _{de}}}})^{M + 1}}\\
&\quad+ \sum\limits_{m = 1}^M {\frac{1}{M}\prod\limits_{i = 1,i \ne m}^M
{(\frac{1}{{2{\alpha _{si}}}} + \frac{1}{{2{\alpha _{id}}}})}
\frac{{M!\alpha _{me}^M}}{{{\alpha _{md}}}}}  \cdot {(\frac{1}{{{\lambda _{de}}}})^M}
\end{split}
\end{equation}
\end{figure*}
Ignoring the higher-order terms in Eq. (65), we have
\begin{equation}\label{equa66}
\begin{split}
P_{{\textrm{intercept}}}^{{\textrm{T-AF{\it{b}}ORS}}} \ge& \sum\limits_{m = 1}^M {\frac{1}{M}
\prod\limits_{i = 1,i \ne m}^M {(\frac{1}{{2{\alpha _{si}}}} + \frac{1}{{2{\alpha _{id}}}})}
\frac{{M!\alpha _{me}^M}}{{{\alpha _{md}}}}}  \\
&\quad\quad \times {(\frac{1}{{{\lambda _{de}}}})^M},
\end{split}
\end{equation}
Substituting Eq. (66) into Eq. (60) gives
\begin{equation}\label{equa67}
{d_{{\textrm{T-AF{\it{b}}ORS}}}} \le M.
\end{equation}
In addition, considering inequalities $\frac{1}{{|{h_{si}}{|^2}}} +
\frac{1}{{|{h_{id}}{|^2}}} \le 2\max
(\frac{1}{{|{h_{si}}{|^2}}},\frac{1}{{|{h_{id}}{|^2}}})$ and
$\frac{1}{x} + \frac{1}{y} \ge \max (\frac{1}{x},\frac{1}{y})$, we
obtain an upper bound on the intercept probability of
T-AF{\it{b}}ORS scheme as
\begin{equation}\label{equa68}
\begin{split}
P_{{\textrm{intercept}}}^{{\textrm{T-AF{\it{b}}ORS}}} \le & \sum\limits_{m = 1}^M
 {\frac{1}{M}\int_0^\infty  {\int_0^\infty { \prod\limits_{i = 1,i \ne m}^M  h_i (x,y)}}}\\
 &\quad \quad \times {[1 - \exp ( - \frac{y}{{\sigma _{md}^2 }})]f(x,y) }  dxdy,
\end{split}
\end{equation}
where $h_i(x,y)=1 - \exp [( - \frac{2}{{\sigma _{si}^2}} -
\frac{2}{{\sigma _{id}^2}})\min (x,y)]$. Similarly to Eq. (66), we
can obtain
\begin{equation}\label{equa69}
\begin{split}
P_{{\textrm{intercept}}}^{{\textrm{T-AF{\it{b}}ORS}}}  \le &\sum\limits_{m = 1}^M
{\frac{1}{M}\prod\limits_{i = 1,i \ne m}^M {(\frac{2}{{{\alpha _{si}}}} +
\frac{2}{{{\alpha _{id}}}})} \frac{{M!\alpha _{me}^M}}{{{\alpha _{md}}}}}\\
&\quad\quad \times {(\frac{1}{{{\lambda _{de}}}})^M},
\end{split}
\end{equation}
for ${\lambda _{de}} \to \infty $. Substituting Eq. (69) into Eq. (60) gives
\begin{equation}\label{equa70}
{d_{{\textrm{T-AF{\it{b}}ORS}}}}  \ge M.
\end{equation}
Therefore, by combining Eqs. (67) and (70), the diversity order of
T-AF{\it{b}}ORS scheme is readily obtained as
\begin{equation}\label{equa71}
{d_{{\textrm{T-AF{\it{b}}ORS}}}} =M,
\end{equation}
which shows that the diversity order $M$ is achieved by T-AF{\it{b}}ORS scheme.
\subsection{T-DFbORS}
In this subsection, we present the diversity order analysis of
T-DF{\it{b}}ORS scheme. Using Eq. (52), we obtain the diversity
order of T-DF{\it{b}}ORS scheme as
\begin{equation}\label{equa72}
{d_{{\textrm{T-DF{\it{b}}ORS}}}} = -\mathop {\lim }\limits_{{\lambda
_{de}} \to \infty } \dfrac{{\log
(P_{{\textrm{intercept}}}^{{\textrm{T-DF{\it{b}}ORS}}})}} {{\log
({\lambda _{de}})}},
\end{equation}
where $P_{{\textrm{intercept}}}^{{\textrm{T-DF{\it{b}}ORS}}}$ is
given in Eq. (45). From Proposition 1, we can similarly obtain $1 -
\exp ( - \frac{x}{{\sigma _{si}^2}} - \frac{x}{{\sigma _{id}^2}}) =
\frac{x}{{\sigma _{si}^2}} + \frac{x}{{\sigma _{id}^2}}$ for
${\lambda _{de}} \to \infty $ by using the Taylor series expansion
and ignoring higher-order terms, from which
$P_{{\textrm{intercept}}}^{{\textrm{T-DF{\it{b}}ORS}}}$ can be
obtained as
\begin{equation}\label{equa73}
\begin{split}
P_{{\textrm{intercept}}}^{{\textrm{T-DF{\it{b}}ORS}}}&=\sum\limits_{m = 1}^M
 {(M - 1)!\prod\limits_{i = 1}^M {(\frac{{{\alpha _{me}}}}{{{\alpha _{si}}}}
 + \frac{{{\alpha _{me}}}}{{{\alpha _{id}}}})} }  \\
 &\quad \quad \quad \times {(\frac{1}{{{\lambda _{de}}}})^M},
\end{split}
\end{equation}
where ${\alpha _{si}} = \sigma _{si}^2/\sigma _{sd}^2$, ${\alpha
_{id}} = \sigma _{id}^2/\sigma _{sd}^2$, and ${\alpha _{me}} =
\sigma _{me}^2/\sigma _{se}^2$. Substituting Eq. (73) into Eq. (72)
yields
\begin{equation}\label{equa74}
{d_{{\textrm{T-DF{\it{b}}ORS}}}} =M,
\end{equation}
which shows that the T-DF{\it{b}}ORS scheme also achieves the
diversity order $M$. As shown in Eqs. (56), (59), (71) and (74), the
P-AF{\it{b}}ORS, P-DF{\it{b}}ORS, T-AF{\it{b}}ORS and
T-DF{\it{b}}ORS schemes all achieve the same diversity order $M$.
This implies that in high MER regions, the intercept probabilities
of P-AF{\it{b}}ORS, P-DF{\it{b}}ORS, T-AF{\it{b}}ORS and
T-DF{\it{b}}ORS schemes all behave as $(1/\lambda_{de})^{M}$ for
$\lambda_{de} \to \infty$. Therefore, for $M>1$, the intercept
probabilities of P-AF{\it{b}}ORS, P-DF{\it{b}}ORS, T-AF{\it{b}}ORS
and T-DF{\it{b}}ORS schemes are reduced much faster than that of
direct transmission as $\lambda_{de} \to \infty$, showing the
physical-layer security benefit of using the optimal relay
selection.

\section{Numerical Results and Discussions}
This section presents the numerical intercept probability results of
conventional direct transmission, T-AF{\it{b}}ORS, T-DF{\it{b}}ORS,
T-AF{\it{b}}MRC, T-DF{\it{b}}MRC, P-AF{\it{b}}ORS and
P-DF{\it{b}}ORS schemes. We show that for both AF and DF protocols,
the proposed optimal relay selection outperforms the traditional
relay selection and multiple relay combining approaches in terms of
intercept probability. Moreover, numerical results also illustrate
that as the number of relays increases, the intercept probabilities
of P-AF{\it{b}}ORS and P-DF{\it{b}}ORS schemes significantly
decrease, showing the security improvement by exploiting cooperative
relays.

\begin{figure}
  \centering
  {\includegraphics[scale=0.6]{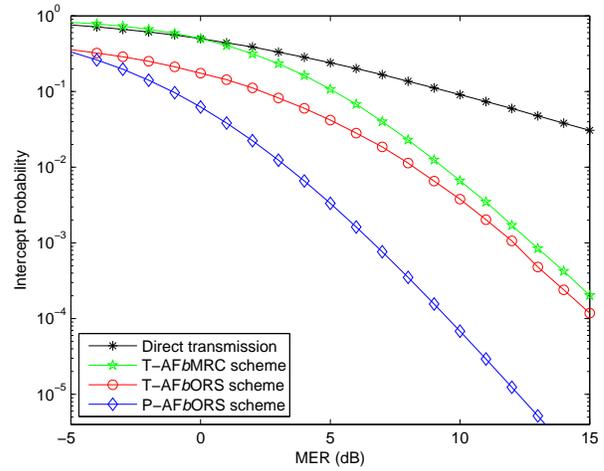}\\
  \caption{{Intercept probability versus main-to-eavesdropper ratio (MER) of
  the direct transmission, T-AF\emph{b}ORS, T-AF\emph{b}MRC, and P-AF\emph{b}ORS
  schemes with $\alpha_{si}=\alpha_{id}=\alpha_{ie}=1$.}}\label{Fig2}}
\end{figure}
{Fig. 2 shows the intercept probability comparison among the direct
transmission, P-AF\emph{b}ORS, T-AF\emph{b}ORS, and T-AF\emph{b}MRC
schemes by plotting Eqs. (32), (34), (42) and (47) as a function of
MER. It is shown from Fig. 2 that the T-AF\emph{b}ORS,
T-AF\emph{b}MRC, and P-AF\emph{b}ORS schemes all perform better than
the direct transmission in terms of intercept probability, implying
the security benefits of exploiting cooperative relays to defend
against eavesdropping attack. One can also see from Fig. 2 that the
intercept probability performance of T-AF\emph{b}MRC scheme is worse
than that of T-AF\emph{b}ORC scheme which performs worse than the
P-AF\emph{b}MRC scheme, showing the advantage of proposed optimal
relay selection over both the traditional relay selection and
multiple relay combining approaches. }
\begin{figure}
  \centering
  {\includegraphics[scale=0.6]{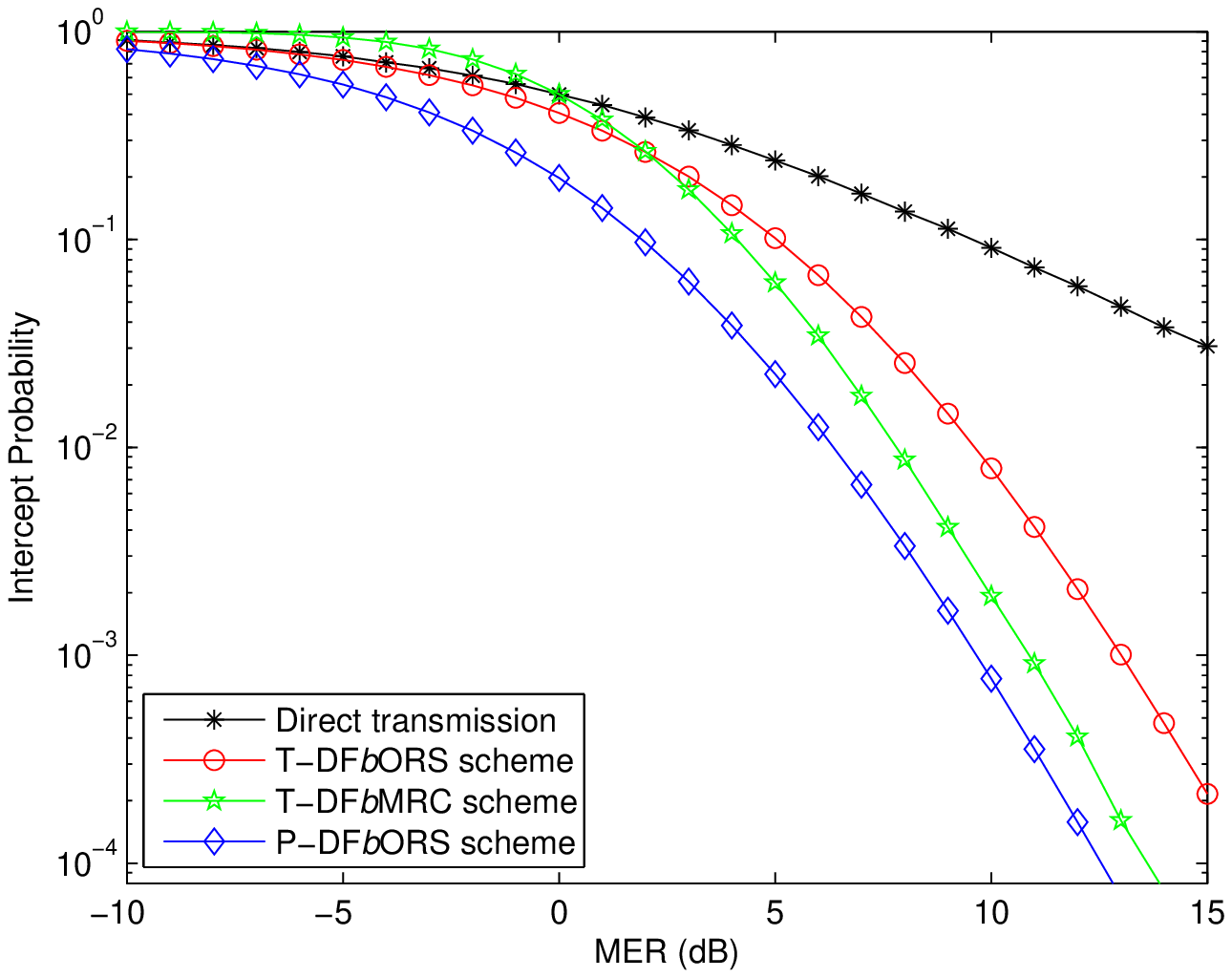}\\
  \caption{{Intercept probability versus main-to-eavesdropper ratio (MER)
  of the direct transmission, T-DF\emph{b}ORS, T-DF\emph{b}MRC, and P-DF\emph{b}ORS
  schemes with ${\rm{Pou}}{{\rm{t}}_i}=10^{-3}$ and
   $\alpha_{si}=\alpha_{id}=\alpha_{ie}=1$.}}\label{Fig3}}
\end{figure}

{In Fig. 3, we show the numerical intercept probability results of
various DF based optimal relay selection and multiple relay
combining schemes, in which the intercept probability curves of
direct transmission, P-DF\emph{b}ORS, T-DF\emph{b}ORS, and
T-DF\emph{b}MRC schemes are plotted by using Eqs. (32), (38), (45),
and (49) with ${\rm{Pou}}{{\rm{t}}_i}=10^{-3}$ and
$\alpha_{si}=\alpha_{id}=\alpha_{ie}=1$. Fig. 3 shows that the
intercept probability of P-DF\emph{b}ORS scheme is always smaller
than that of T-DF\emph{b}MRC scheme which further outperforms the
T-DF\emph{b}ORS scheme in terms of intercept probability. In other
words, the P-DF\emph{b}ORS scheme achieves the best intercept
probability performance, further confirming the advantage of
proposed optimal relay selection over traditional relay selection
and multiple relay combining. Therefore, no matter which relaying
protocol (i.e., AF and DF) is considered, the proposed optimal relay
selection always performs better than the traditional relay
selection and multiple relay combining approaches in terms of
intercept probability. }
\begin{figure}
  \centering
  {\includegraphics[scale=0.6]{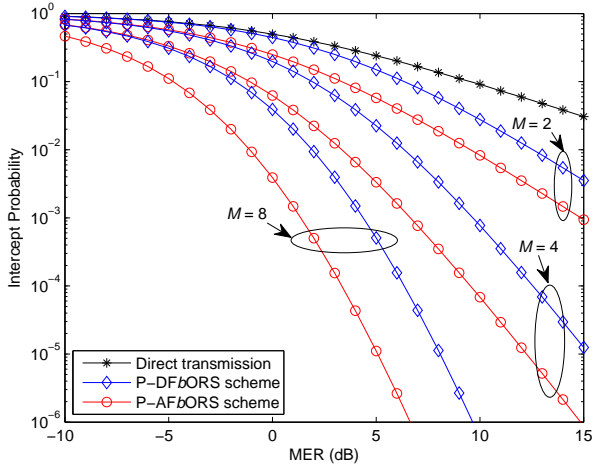}\\
  \caption{Intercept probability versus main-to-eavesdropper ratio (MER) of
  the direct transmission, P-AF\emph{b}ORS, and P-DF\emph{b}ORS schemes with
   $\alpha_{si}=\alpha_{id}=\alpha_{ie}=1$.}\label{Fig4}}
\end{figure}

Fig. 4 depicts the intercept probability comparison between the
P-AF\emph{b}ORS and P-DF\emph{b}ORS schemes with
$\alpha_{si}=\alpha_{id}=\alpha_{ie}=1$. One can see from Fig. 4
that for the cases of $M=2$, $M=4$, and $M=8$, the direct
transmission strictly performs worse than the P-AF\emph{b}ORS and
P-DF\emph{b}ORS schemes in terms of intercept probability. Moreover,
as the number of relays $M$ increases from $M=2$ to $M=8$, the
intercept probabilities of P-AF\emph{b}ORS and P-DF\emph{b}ORS
schemes both decrease significantly. This means that increasing the
number of cooperative relays can enhance the physical-layer security
against eavesdropping attack. In addition, Fig. 4 also shows that
for the cases of $M=2$, $M=4$, and $M=8$, the P-AF\emph{b}ORS scheme
always outperforms the P-DF\emph{b}ORS scheme, showing the advantage
of AF relaying protocol over DF protocol.

\begin{figure}
  \centering
  {\includegraphics[scale=0.6]{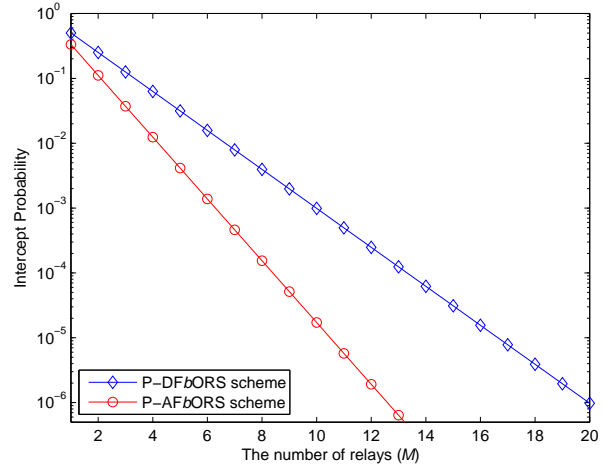}\\
  \caption{Intercept probability versus the number of relays $M$ of the P-AF\emph{b}ORS
   and P-DF\emph{b}ORS schemes with $\lambda_{de}=3{\textrm{dB}}$ and
   $\alpha_{si}=\alpha_{id}=\alpha_{ie}=1$.}\label{Fig5}}
\end{figure}

Fig. 5 shows the intercept probability versus the number of relays
$M$ of the P-AF\emph{b}ORS and P-DF\emph{b}ORS schemes with
$\lambda_{de}=3{\textrm{dB}}$ and
$\alpha_{si}=\alpha_{id}=\alpha_{ie}=1$. It is observed from Fig. 5
that the P-AF\emph{b}ORS scheme strictly performs better the
P-DF\emph{b}ORS scheme in terms of intercept probability. One can
also see from Fig. 5 that as the number of relays $M$ increases, the
intercept probabilities of both P-AF\emph{b}ORS and P-DF\emph{b}ORS
schemes significantly decrease, showing the wireless security
improvement with an increasing number of relays. In addition, as
shown in Fig. 5, the intercept probability improvement of
P-AF\emph{b}ORS over P-DF\emph{b}ORS becomes more significant as the
number of relays increases.

\section{Conclusion}
In this paper, we explored the relay selection for improving
physical-layer security in cooperative wireless networks and
proposed the AF and DF based optimal relay selection schemes, i.e.,
P-AF\emph{b}ORS and P-DF\emph{b}ORS. For the purpose of comparison,
we also examined the conventional direct transmission,
T-AF\emph{b}ORS, T-DF\emph{b}ORS, T-AF\emph{b}MRC, and
T-DF\emph{b}MRC schemes. We derived closed-form intercept
probability expressions of the direct transmission, T-AF\emph{b}ORS,
T-DF\emph{b}ORS, T-AF\emph{b}MRC, T-DF\emph{b}MRC, P-AF\emph{b}ORS
and P-DF\emph{b}ORS schemes over Rayleigh fading channels. We
further analyzed the diversity order performance of the traditional
and proposed optimal relay selection schemes and showed that for
both AF and DF protocols, the proposed and traditional relay
selection schemes achieve the diversity order $M$, where $M$ is the
number of cooperative relays. Numerical results also illustrated
that no matter which relaying protocol is considered (i.e., AF and
DF), the proposed optimal relay selection strictly outperforms the
traditional relay selection and multiple relay combining approaches
in terms of intercept probability. In addition, as the number of
relays increases, the intercept probability performance of both
P-AF\emph{b}ORS and P-DF\emph{b}ORS significantly improves, implying
the wireless security enhancement with an increasing number of
cooperative relays.

It is worth mentioning that we only investigated the single-source
and single-destination for cooperative relay networks in this paper.
In future, we will extend the results of this paper to a general
case with multiple-source and multiple-destination, for which the
opportunistic transmission scheduling may be exploited to defend
against eavesdropping attack. More specifically, a source node with
the highest secrecy capacity can be opportunistically scheduled to
transmit to its destination. Once a source-destination pair is
determined with the transmission scheduling policy, we can consider
the use of optimal relay selection developed in this paper to assist
the transmission between source and destination against the
eavesdropping attack.

\appendices
\section{Proof of Proposition 1}
Denoting $z = (\frac{1}{{2\sigma _{si}^2}} + \frac{1}{{2\sigma
_{id}^2}})\min (x,y)$ and using the joint PDF of $(X,Y)$ in Eq.
(62), we can obtain the mean of $z$ as
\begin{equation}
\begin{split}
E(z) &= (\frac{1}{{2{\alpha _{si}}}} + \frac{1}{{2{\alpha _{id}}}})\frac{{{\alpha _{sm}}
\alpha _{me}^2}}{{{{({\alpha _{sm}} + {\alpha _{me}}\lambda _{de}^{ - 1})}^2}}} \cdot \frac{1}
{{\lambda _{de}^2}}\\
&\quad+(\frac{1}{{2{\alpha _{si}}}} + \frac{1}{{2{\alpha _{id}}}})\frac{{\alpha _{sm}^2
{\alpha _{me}}}}{{{{({\alpha _{sm}} + {\alpha _{me}}\lambda _{de}^{ - 1})}^2}}}
\cdot \frac{1}{{{\lambda _{de}}}},
\end{split}\label{A.1}\tag{A.1}
\end{equation}
where ${\alpha _{si}} = \sigma _{si}^2/\sigma _{sd}^2$, ${\alpha
_{id}} = \sigma _{id}^2/\sigma _{sd}^2$, ${\alpha _{sm}} = \sigma
_{sm}^2/\sigma _{sd}^2$, and ${\alpha _{me}} = \sigma _{me}^2/\sigma
_{se}^2$. Considering ${\lambda _{de}} \to \infty $ and ignoring the
higher-order term, we have
\begin{equation} E(z) = (\frac{{{\alpha
_{me}}}}{{2{\alpha _{si}}}} + \frac{{{\alpha _{me}}}}{{2{\alpha
_{id}}}}) \cdot \frac{1}{{{\lambda _{de}}}},\label{A.2}\tag{A.2}
\end{equation}
which shows that $E(z)$ converges to zero as ${\lambda _{de}} \to
\infty $. Moreover, using Eq. (62) and letting ${\lambda _{de}} \to
\infty $, we can obtain $E({z^2})$ as
\begin{equation}
\begin{split}
E({z^2}) = {(\frac{{{\alpha _{me}}}}{{2{\alpha _{si}}}} + \frac{{{\alpha _{me}}}}
{{2{\alpha _{si}}}})^2} \cdot \frac{2}{{\lambda _{de}^2}},
\end{split}\label{A.3}\tag{A.3}
\end{equation}
where the third equation is obtained by ignoring higher-order terms.
From Eqs. (A.2) and (A.3), the variance of $z$ is given by
\begin{equation}
Var(z) = E({z^2}) - {[E(z)]^2} = {(\frac{{{\alpha _{me}}}}{{2{\alpha _{si}}}}
+ \frac{{{\alpha _{me}}}}{{2{\alpha _{si}}}})^2} \cdot \frac{1}{{\lambda _{de}^2}},\label{A.4}\tag{A.4}
\end{equation}
for ${\lambda _{de}} \to \infty $. It is shown from Eqs. (A.2) and
(A.4) that both mean and variance of $z$ approach to zero as
${\lambda _{de}} \to \infty $, implying that $z \to 0$ as ${\lambda
_{de}} \to \infty $. Thus, considering ${\lambda _{de}} \to \infty $
and using Taylor series expansion, we obtain
\begin{equation}
1 - \exp ( - z) = z + O(z),\label{A.5}\tag{A.5}
\end{equation}
where $O(z)$ represents higher-order infinitesimal. Substituting $z
= (\frac{1}{{2\sigma _{si}^2}} + \frac{1}{{2\sigma _{id}^2}})\min
(x,y)$ into Eq. (A.5) and ignoring higher-order infinitesimal, we
have
\begin{equation}
\begin{split}
&1 - \exp [( - \frac{1}{{2\sigma _{si}^2}} -
\frac{1}{{2\sigma _{id}^2}})\min (x,y)] \\
&= (\frac{1}{{2\sigma_{si}^2}} + \frac{1}{{2\sigma _{id}^2}})\min
(x,y).
\end{split}\label{A.6}\tag{A.6}
\end{equation}
In addition, denoting $t = \frac{y}{{\sigma _{md}^2}}$, we can
easily obtain both mean and variance of $t$ as
\begin{equation} E(t)
= \int_0^\infty  {\frac{y}{{\sigma _{md}^2\sigma _{me}^2}}\exp ( -
\frac{y}{{\sigma _{me}^2}})dy = } \frac{{{\alpha _{me}}}}{{{\alpha
_{md}}}} \cdot \frac{1}{{{\lambda _{de}}}},\label{A.7}\tag{A.7}
\end{equation}
and
\begin{equation}
Var(t) = E({t^2}) - {[E(t)]^2} = \frac{{\alpha _{me}^2}}{{\alpha _{md}^2}}
\cdot \frac{1}{{\lambda _{de}^2}}.\label{A.8}\tag{A.8}
\end{equation}
One can observe from Eqs. (A.7) and (A.8) that both mean and
variance of $t$ approach to zero as ${\lambda _{de}} \to \infty $,
meaning that $t \to 0$ as ${\lambda _{de}} \to \infty $. Hence,
considering ${\lambda _{de}} \to \infty $ and using Taylor series
expansion, we obtain
\begin{equation} 1 - \exp ( - t) = t + O(t).
\label{A.9}\tag{A.9}
\end{equation}
Substituting $t = \frac{y}{{\sigma _{md}^2}}$ into Eq. (A.9) and
ignoring the higher-order infinitesimal yield
\begin{equation} 1 -
\exp ( - \frac{y}{{\sigma _{md}^2}}) = \frac{y}{{\sigma
_{md}^2}},\label{A.10}\tag{A.10}
\end{equation}
which completes the proof of Proposition 1.

\begin{IEEEbiography}[{\includegraphics[width=1in,height=1.25in]{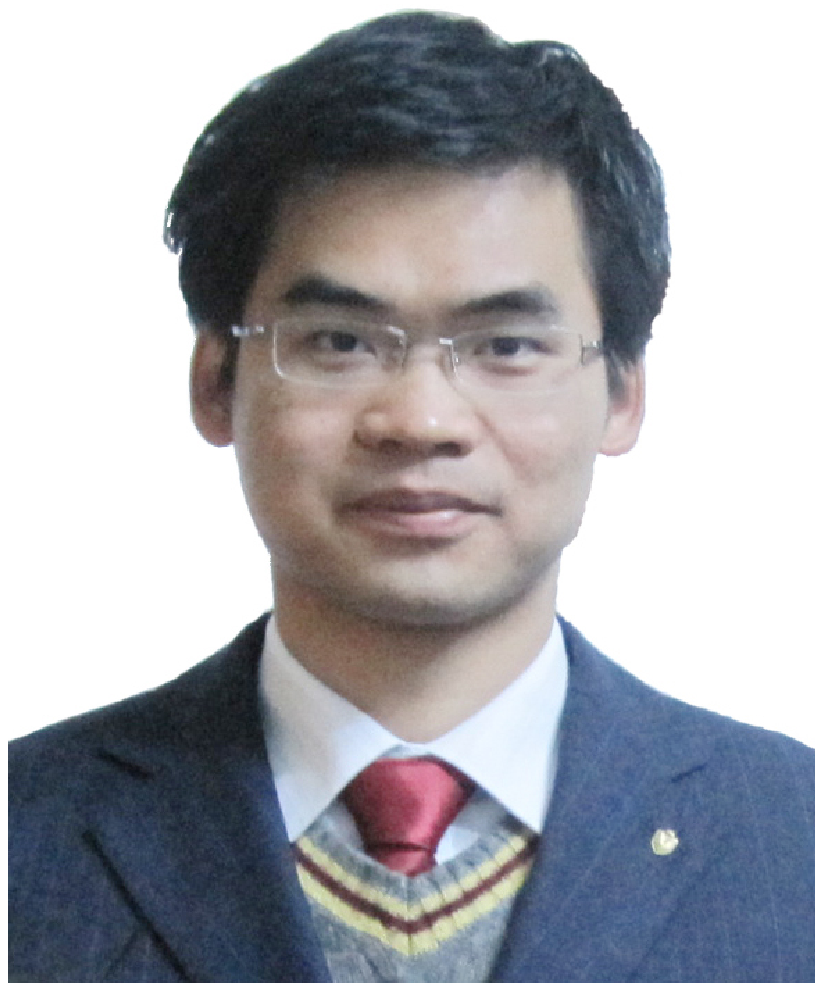}}]{Yulong Zou}
received the B.Eng. degree in information engineering from the
Nanjing University of Posts and Telecommunications (NUPT), Nanjing,
China, in July 2006. Then, he was admitted to the Graduate School of
NUPT to pursue his Ph.D. degree in September 2006. In May 2009, he
was invited to serve as a visiting scholar at the Stevens Institute
of Technology (Stevens Tech), New Jersey, United States. From
September 2009, he joined a new Ph.D. program in electrical
engineering at Stevens Tech under the full sponsorship of United
States Department of Defense and started working toward his two
Ph.D. degrees concurrently at NUPT and Stevens Tech, respectively.
He received the first Ph.D. degree from the Stevens Institute of
Technology in May 2012 and the second Ph.D. degree from the Nanjing
University of Posts and Telecommunications in July 2012.

Dr. Zou is currently serving as an editor for the IEEE
Communications Surveys \& Tutorials, IEEE Communications Letters,
EURASIP Journal on Advances in Signal Processing, and KSII
Transactions on Internet and Information Systems. He is also serving
as a lead guest editor for special issue on ``Security Challenges
and Issues in Cognitive Radio Networks" in the EURASIP Journal on
Advances in Signal Processing. In addition, he has acted as
symposium chairs, session chairs, and TPC members for a number of
IEEE sponsored conferences including the IEEE Wireless
Communications and Networking Conference (WCNC), IEEE Global
Telecommunications Conference (GLOBECOM), IEEE International
Conference on Communications (ICC), IEEE Vehicular Technology
Conference (VTC), International Conference on Communications in
China (ICCC), International Conference on Communications and
Networking in China (ChinaCom), and so on.

His research interests span a wide range of topics in wireless
communication and signal processing including the cooperative
communications, cognitive radio, wireless security, and green
communications. In these areas, he has published extensively in
internationally renowned journals including the IEEE Transactions on
Signal Processing, IEEE Transactions on Communications, IEEE Journal
on Selected Areas in Communications, IEEE Transactions on Wireless
Communications, and IEEE Communications Magazine.

\end{IEEEbiography}

\begin{IEEEbiography}[{\includegraphics[width=1in,height=1.25in]{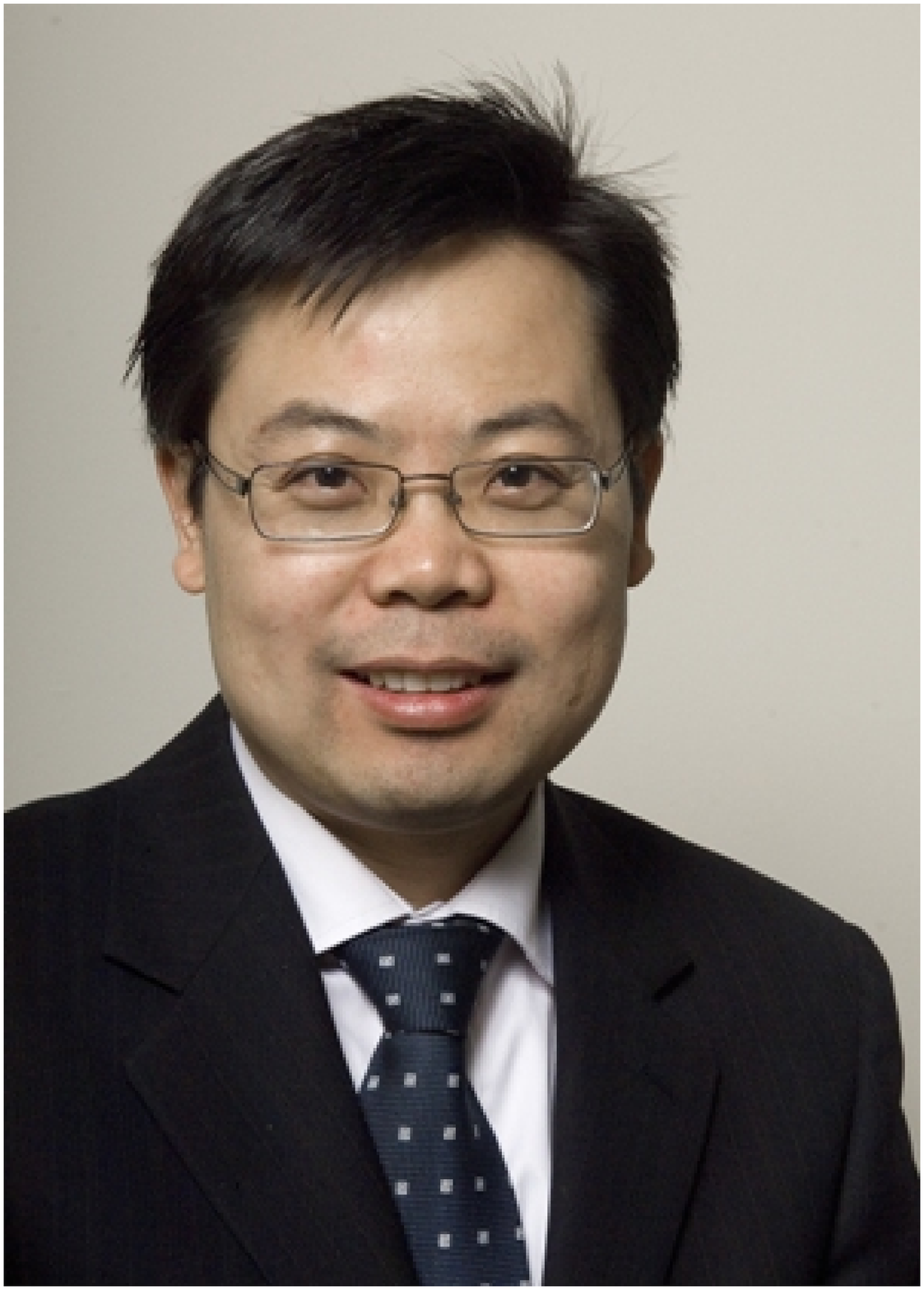}}]
{Xianbin Wang}(S'98-M'99-SM'06) is an Associate Professor at The
University of Western Ontario and a Canada Research Chair in
Wireless Communications. He received his Ph.D. degree in electrical
and computer engineering from National University of Singapore in
2001.

Prior to joining Western, he was with Communications Research Centre
Canada as Research Scientist/Senior Research Scientist between July
2002 and Dec. 2007. From Jan. 2001 to July 2002, he was a system
designer at STMicroelectronics, where he was responsible for system
design for DSL and Gigabit Ethernet chipsets. He was with Institute
for Infocomm Research, Singapore (formerly known as Centre for
Wireless Communications), as a Senior R \& D engineer in 2000. His
primary research area is wireless communications and related
applications, including adaptive communications, wireless security,
and wireless infrastructure based position location. Dr. Wang has
over 150 peer-reviewed journal and conference papers on various
communication system design issues, in addition to 23 granted and
pending patents and several standard contributions.

Dr. Wang is an IEEE Distinguished Lecturer and a Senior Member of
IEEE. He was the recipient of three IEEE Best Paper Awards. He
currently serves as an Associate Editor for IEEE Wireless
Communications Letters, IEEE Transactions on Vehicular Technology
and IEEE Transactions on Broadcasting. He was also an editor for
IEEE Transactions on Wireless Communications between 2007 and 2011.
Dr. Wang was involved in a number of IEEE conferences including
GLOBECOM, ICC, WCNC, VTC, and ICME, on different roles such as
symposium chair, track chair, TPC and session chair.

\end{IEEEbiography}

\begin{IEEEbiography}[{\includegraphics[width=1in,height=1.25in]{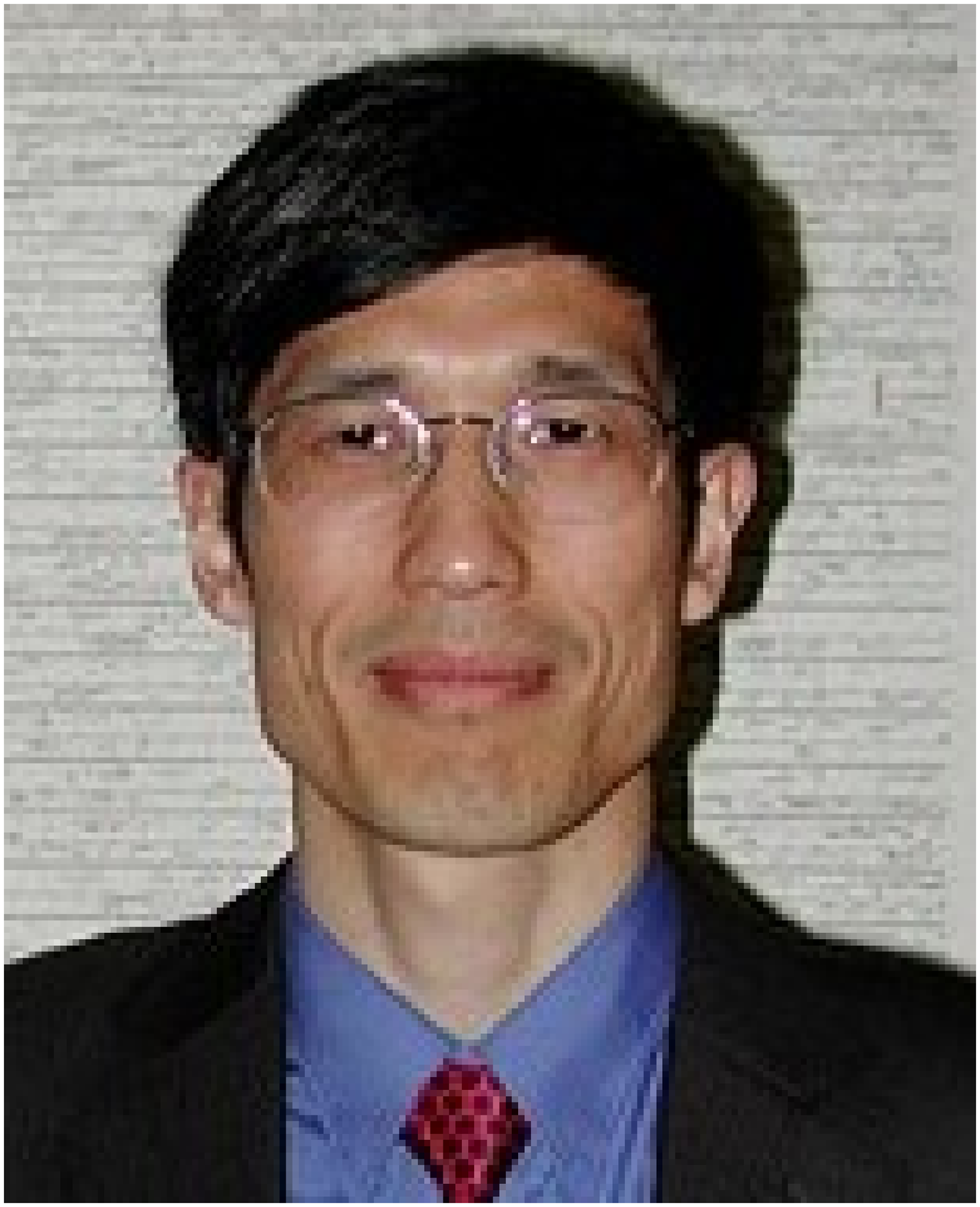}}]{Weiming Shen}
is a Senior Research Scientist at the National Research Council
Canada and an Adjunct Research Professor at the University of
Western Ontario. He is a Fellow of IEEE. He received his Bachelor
and Master¡¯s degrees from Northern (Beijing) Jiaotong University,
China and his PhD degree from the University of Technology of
Compi¨¨gne, France. His recent research interest includes
agent-based collaboration technology and applications, wireless
sensor networks. He has published several books and over 300 papers
in scientific journals and international conferences in the related
areas. His work has been cited over 6,000 times with an h-index of
37. He has been invited to provide over 60 invited lectures/seminars
at different academic and research institutions over the world and
keynote presentations / tutorials at various international
conferences. He is a member of the Steering Committee for the IEEE
Transactions on Affective Computing and an Associate Editor or
Editorial Board Member of ten international journals (including IEEE
Transactions on Automation Science and Engineering, Computers in
Industry; Advanced Engineering Informatics; Service Oriented
Computing and Applications) and served as guest editor for several
other international journals.
\end{IEEEbiography}

\end{document}